\documentclass{elsart}
\usepackage{amsmath,amssymb}
\usepackage{graphicx}
\usepackage{subfigure}
\usepackage{hyperref}

\begin{document}

\begin{frontmatter}

\title{$\eta$ and $\eta'$ mesons from $N_f=2+1+1$ twisted mass
  lattice QCD}

\begin{center}
  \includegraphics[draft=false]{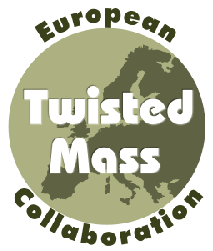}
\end{center}

\author{ETM Collaboration},
\author[a]{Konstantin Ottnad},
\author[b]{Chris Michael},
\author[c]{Siebren Reker},
\author[a]{Carsten Urbach}


\address[a]{Helmholtz Institut f{\"u}r Strahlen- und Kernphysik,
  Universit{\"a}t Bonn\\
Nussallee 14-16, 53115 Bonn, Germany}
\address[b]{Theoretical Physics Division, Department of Mathematical
  Sciences,\\
  The University of Liverpool, Liverpool L69 3BX, UK}
\address[c]{Centre for Theoretical Physics, University of Groningen,\\
  Nijenborgh 4, 9747 AG Groningen, The Netherlands}

\begin{abstract}
  We determine mass and mixing angles of $\eta$ and $\eta'$
  states using $N_f=2+1+1$ Wilson twisted mass lattice QCD. We
  describe how those flavour singlet states need to be treated in this
  lattice formulation. Results are presented for three values of 
  the lattice spacing, $a=0.061\ \mathrm{fm}$, $a=0.078\
  \mathrm{fm}$ and $a=0.086\ \mathrm{fm}$, with light
  quark masses corresponding to values of the charged pion mass in a
  range of $230$ to $500\ \mathrm{MeV}$ and fixed bare strange and
  charm quark mass values. We obtain $M_\eta=557(15)(45)\ \mathrm{MeV}$
  (first error statistical, second systematic) and 
  $\phi=44(5)^{\circ}$ for a single mixing angle in the
  quark flavour basis, $\theta=-10(5)^\circ$ in the octet-singlet
  basis.
\end{abstract}

\end{frontmatter}

\section{Introduction}

From experiments it is known that the masses of the nine light
pseudo-scalar
mesons show an interesting pattern. Taking the quark model point of
view, the three lightest mesons, the pions, contain only the two
lightest quark flavours, the \emph{up}- and \emph{down}-quarks. The
pion triplet has a mass of $M_\pi\approx140\ \mathrm{MeV}$. For the other six,
the \emph{strange} quark also contributes, and hence they are
heavier. In contrast to what one might expect, five of them, 
the four kaons and the $\eta$ meson,  have
roughly equal mass around $500$ to $600\ \mathrm{MeV}$, while
the last one, the $\eta'$ meson, is much
heavier, with a mass of about $1\ \mathrm{GeV}$. On the QCD level, the
reason for this pattern is thought to be the breaking of the $U_A(1)$
symmetry by quantum effects. The $\eta'$ meson is, even in a
world with three massless quarks, not a Goldstone boson.

In this paper, we present the first lattice study of $\eta$
and $\eta'$ meson masses using mass degenerate \emph{up}, \emph{down}
as well as heavier, non-degenerate \emph{strange} and \emph{charm} dynamical
quark flavours. The lattice QCD formulation is the Wilson twisted mass
formulation~\cite{Frezzotti:2000nk} with 
$N_f=2+1+1$ dynamical quark flavours. This will not only allow a study
of the dependence of the $\eta,\eta'$ masses on the light quark mass
value, but also an investigation of the charm quark contribution to 
both of these states. Moreover, the $\eta_c$ meson could be studied
in the unitary case in principle. 

$\eta$ and $\eta'$ states are difficult to treat in lattice QCD,
because fermionic disconnected contributions appear and cannot be
ignored. This is why the amount of available results for these
states from lattice QCD is rather limited in range of pion mass values
as well as values of the lattice spacing. For recent lattice studies
in $N_f=2+1$ flavour QCD
see~\cite{Christ:2010dd,Kaneko:2009za,Dudek:2011tt,Gregory:2011sg}.
Some of these we will discuss in more detail later and compare with
our results. A previous lattice study with Wilson twisted mass
fermions has been performed with $N_f=2$ dynamical quark
flavours~\cite{Jansen:2008wv}. In this work we have developed
particular noise reduction techniques, which we also used to obtain the
results presented here for the $N_f=2+1+1$ case.

The paper is organised as follows. In section \ref{sec:intro} we
introduce the lattice QCD framework we are using, the Wilson twisted
mass formalism, followed by a discussion of how we deal with flavour
singlet pseudo-scalar mesons in this framework in section
\ref{sec:tmfs}. In section \ref{sec:res} we present our results and
discuss them in the following final section. More details on our
analysis procedure and data tables can be found in the appendix.

\begin{table}[t!]
  \centering
  \begin{tabular*}{1.\textwidth}{@{\extracolsep{\fill}}lcccccccc}
    \hline\hline
    ensemble & $\beta$ & $a\mu_\ell$ & $a\mu_\sigma$ & $a\mu_\delta$ &
    $L/a$ & $N_\mathrm{conf}$ & $N_s$ & $N_b$ \\ 
    \hline\hline
    A30.32   & $1.90$ & $0.0030$ & $0.150$  & $0.190$  & $32$ & $1367$ & $24$ & $5$ \\
    A40.24   & $1.90$ & $0.0040$ & $0.150$  & $0.190$  & $24$ & $2630$ & $32$ & $10$ \\
    A40.32   & $1.90$ & $0.0040$ & $0.150$  & $0.190$  & $32$ & $863$  & $24$ & $4$ \\
    A60.24   & $1.90$ & $0.0060$ & $0.150$  & $0.190$  & $24$ & $1251$ & $32$ & $5$ \\
    A80.24   & $1.90$ & $0.0080$ & $0.150$  & $0.190$  & $24$ & $2449$ & $32$ & $10$ \\
    A100.24  & $1.90$ & $0.0100$ & $0.150$  & $0.190$  & $24$ & $2493$ & $32$ & $10$ \\
    \hline
    A80.24s  & $1.90$ & $0.0080$ & $0.150$  & $0.197$  & $24$ & $2517$ & $32$ & $10$ \\
    A100.24s & $1.90$ & $0.0100$ & $0.150$  & $0.197$  & $24$ & $2312$ & $32$ & $10$ \\
    \hline
    B25.32   & $1.95$ & $0.0025$ & $0.135$  & $0.170$  & $32$ & $1484$ & $24$ & $5$ \\
    B35.32   & $1.95$ & $0.0035$ & $0.135$  & $0.170$  & $32$ & $1251$ & $24$ & $5$ \\
    B55.32   & $1.95$ & $0.0055$ & $0.135$  & $0.170$  & $32$ & $1545$ & $24$ & $5$ \\ 
    B75.32   & $1.95$ & $0.0075$ & $0.135$  & $0.170$  & $32$ & $922$  & $24$ & $4$ \\ 
    B85.24   & $1.95$ & $0.0085$ & $0.135$  & $0.170$  & $24$ & $573$  & $32$ & $2$ \\
    \hline
    D15.48   & $2.10$ & $0.0015$ & $0.120$  & $0.1385$ & $48$ & $1045$ & $24$ & $10$ \\
    D45.32sc & $2.10$ & $0.0045$ & $0.0937$ & $0.1077$ & $32$ & $1887$ & $24$ & $10$ \\
    \hline\hline
    \vspace*{0.1cm}
  \end{tabular*}
  \caption{The ensembles used in this investigation. The notation of
    ref.~\cite{Baron:2010bv} is used for labeling the ensembles. In
    addition we give the number of configurations $N_\mathrm{conf}$,
    the number of stochastic samples $N_s$ for all ensembles and the
    bootstrap block length $N_b$.}
  \label{tab:setup}
\end{table}

\section{Lattice Action}
\label{sec:intro}

The lattice QCD action for $N_f=2+1+1$ Wilson twisted mass fermions
reads 
\begin{equation}
  \label{eq:action}
  S = S_g + \bar\chi_\ell\, D_\ell\, \chi_\ell +
  \bar\chi_h\, D_h\, \chi_h\, .
\end{equation}
For the gauge action $S_g$ we use the Iwasaki gauge
action~\cite{Iwasaki:1985we}. The twisted mass Dirac 
operator for the light -- i.e. up/down quark -- doublet
reads~\cite{Frezzotti:2000nk}
\begin{equation}
  \label{eq:Dud}
  D_\ell = D_W + m_0 + i \mu_\ell \gamma_5\tau^3
\end{equation}
and for the strange/charm doublet~\cite{Frezzotti:2003xj}
\begin{equation}
  \label{eq:Dsc}
  D_h = D_W + m_0 + i \mu_\sigma \gamma_5\tau^1 + \mu_\delta \tau^3\, ,
\end{equation}
where $D_W$ is the Wilson Dirac operator. The value of $m_0$ was tuned
to its critical value $m_\mathrm{crit}$ as discussed in
refs.~\cite{Chiarappa:2006ae,Baron:2010bv} in 
order to realise automatic $\mathcal{O}(a)$ improvement at maximal
twist~\cite{Frezzotti:2003ni}. This way of $\mathcal{O}(a)$
improvement was first shown to work in 
practice in refs.~\cite{Jansen:2003ir,Jansen:2005kk} in the quenched
approximation, and later for $N_f=2$ in
ref.~\cite{Baron:2009wt} (see also ref.~\cite{Blossier:2010cr}). For 
a review see ref.~\cite{Shindler:2007vp}. Also
for $N_f=2+1+1$ we have strong indications that the scaling properties
of Wilson twisted mass lattice QCD at maximal twist are
favourable~\cite{Baron:2010bv}. For a discussion on how to determine
kaon and D-meson masses see ref.~\cite{Baron:2010th}.
The bare twisted masses
$\mu_{\sigma}$ and $\mu_{\delta}$ are related to the bare strange and
charm quark masses via the relation
\begin{equation}
  \label{eq:msc}
  m_{c,s} = \mu_\sigma\ \pm\ Z \mu_\delta
\end{equation}
where $Z \equiv (Z_\mathrm{P}/Z_\mathrm{S})$ denotes the ratio of
pseudo-scalar and scalar renormalisation constants $Z_\mathrm{P}$ 
and $Z_\mathrm{S}$. Quark fields in the twisted basis are denoted by
$\chi_{\ell,h}$ and in the physical basis by
$\psi_{\ell,h}$. They are related via the axial rotations
\begin{equation}
  \begin{split}
    \psi_\ell = e^{i\pi\gamma_5\tau^3/4}\chi_\ell\,,& \quad\bar\psi_\ell
    = \bar\chi_\ell\ e^{i\pi\gamma_5\tau^3/4}\,,\\
    \psi_h = e^{i\pi\gamma_5\tau^1/4}\chi_h\,,& \quad\bar\psi_h
    = \bar\chi_h\ e^{i\pi\gamma_5\tau^1/4}\,.\\
  \end{split}
\end{equation}
With automatic $\mathcal{O}(a)$ improvement being the biggest
advantage of twisted mass lattice QCD (tmQCD) at maximal twist, the
downside is that flavour (and parity) 
symmetry is broken at finite values of the lattice spacing. This was
theoretically and numerically shown to affect mainly the mass value of
the neutral pion~\cite{Baron:2009wt,Urbach:2007rt,Dimopoulos:2009qv},
however, in  the case of $N_f=2+1+1$ dynamical quarks, it implies the
complication of mixing between strange and charm quarks.

\begin{table}[t!]
 \centering
 \begin{tabular*}{.8\textwidth}{@{\extracolsep{\fill}}lccc}
  \hline\hline
  $\beta$   & 1.90    & 1.95    & 2.10 \\
  $r_0^\chi / a$ & 5.231(38) & 5.710(41) & 7.538(58) \\
  \hline\hline
  \vspace*{0.1cm}
 \end{tabular*}
 \caption{Values of $r_0^\chi/a$ for the three $\beta$-values.}
 \label{tab:r0values}
\end{table}

We use gauge configurations as produced by the European Twisted Mass
Collaboration (ETMC) with action described
above~\cite{Baron:2010bv,Baron:2010th,Baron:2011sf}. The details of the
configurations are described in ref.~\cite{Baron:2010bv} and the
ensembles used in this investigation are summarised in 
table~\ref{tab:setup}: we use ensembles denoted with $A$, $B$ and $D$
with values of the lattice spacing $a_A=0.0863(4) \mathrm{fm}$, 
$a_B=0.0779(4) \mathrm{fm}$ and $a_D=0.0607(2) \mathrm{fm}$, 
corresponding to $\beta_A=1.90$, $\beta_B=1.95$ and $\beta_D = 2.10$,
respectively~\cite{Baron:2011sf}. The physical volumes are with only
a few exceptions larger than $3\ \mathrm{fm}$.
In the table we also compile the number of investigated gauge
configurations and the number of stochastic samples per gauge
configuration used to estimate the disconnected contributions. 

Throughout this paper we will use the Sommer parameter
$r_0$~\cite{Sommer:1993ce} to investigate the scaling of our
results. We use the values of $r^\chi_0/a$ extrapolated to the
massless limit at each $\beta$-value 
separately. For $\beta=1.90$ and $\beta=1.95$ we use the values quoted
in ref.~\cite{Baron:2010bv}. For $\beta=2.10$ we did the extrapolation
by ourselves. All values for $r_0^\chi/a$ are shown in
table~\ref{tab:r0values}. For setting the physical scale we could use
the results of ref.~\cite{Baron:2011sf}, where a chiral fit to data
for $f_\mathrm{PS}$ and $m_\mathrm{PS}$ and the physical value of
$f_\pi$ was used to set the scale. However, the fit in
ref.~\cite{Baron:2011sf} includes only two 
data points at $\beta=2.10$, and hence, is rather
preliminary. Therefore, we prefer to use a value of $r_0=0.45(2)\
\mathrm{fm}$ in 
this paper to set the scale. The 5\% error covers the statistical
uncertainty and spread quoted in ref.~\cite{Baron:2011sf} and allows
room for systematic uncertainties. As soon as an update of the scale
setting becomes available the results in this paper can be
updated accordingly. For fixing the light and strange quark masses to
their physical values we will use the experimental values of
$M_{\pi^0}=135\ \mathrm{MeV}$ and $M_{\mathrm{K}^0}=498\
  \mathrm{MeV}$. We use the masses of the neutral 
mesons to reduce uncertainties from the fact that we do not include
electromagnetic effects in our simulation. See also
ref.~\cite{Colangelo:2010et} and references therein for a discussion.

For every $\beta$-value the bare values of $a\mu_\sigma$ and
$a\mu_\delta$ were kept fixed, and only the value of the light twisted
mass parameter $a\mu_\ell$ is varied. The kaon masses measured on
these ensembles are close to the physical value with a deviation of
up to $10\%$~\cite{Baron:2010bv} in particular for the $A$ and
$D$ ensembles. The D-meson mass values have a large uncertainty, but
they are also close to physical~\cite{Baron:2010bv}. We will discuss
this point later in more detail.

For the ensembles $A80.24$ and $A100.24$ we have additional ensembles
with re-tuned values of $a\mu_\sigma$ and $a\mu_\delta$ denoted by
$A80.24s$ and $A100.24s$ (see table~\ref{tab:setup}), which reproduce
the physical kaon mass value 
more accurately than the original $A80.24$ and $A100.24$ ensembles
(see figure~\ref{fig:metaA}). We
will use these ensembles to estimate the strange quark mass
dependence of the $\eta$ mass.

\section{Flavour singlet pseudo-scalar mesons }
\label{sec:tmfs}

In order to compute masses of pseudo-scalar flavour singlet mesons we
have to include light, strange and charm contributions to build the
appropriate correlation functions. In the light sector, one appropriate
operator is given by~\cite{Jansen:2008wv}
\begin{equation}
  \label{eq:ll}
  \frac{1}{\sqrt{2}}(\bar{\psi}_u i \gamma_5\psi_u + \bar{\psi}_d i \gamma_5\psi_d)
  \quad \to\quad \frac{1}{\sqrt{2}}(- \bar\chi_u \chi_u + \bar\chi_d
  \chi_d)\ \equiv\ \mathcal{O}_{\ell}\, , 
\end{equation}
in the physical and the twisted basis, respectively. With twisted mass
fermions we have to work with doublets of quarks, hence, in the
strange and charm sector the corresponding operators read 
\begin{equation}
  \label{eq:physsinglet}
  \begin{pmatrix}
    \bar \psi_c \\
    \bar \psi_s \\
  \end{pmatrix}^T
  i \gamma_5 \frac{1 \pm \tau^3}{2}
  \begin{pmatrix}
    \psi_c \\
    \psi_s \\
  \end{pmatrix}
  \quad\to\quad
  \begin{pmatrix}
    \bar \chi_c \\
    \bar \chi_s \\
  \end{pmatrix}^T
  \frac{-\tau^1  \pm  i \gamma_5\tau^3}{2}
  \begin{pmatrix}
    \chi_c \\
    \chi_s \\
  \end{pmatrix}\ \equiv\ \mathcal{O}_{c,s}\, .
\end{equation}
The sign in $1\pm\tau^3$ in the physical basis distinguishes the
charm and strange quark contribution. As a consequence, working in the
twisted basis we need to compute correlation functions of the
following interpolating operators
\begin{equation}
  \label{eq:psequ}
  \begin{split}
    \mathcal{O}_{c}\ &\equiv\
    Z(\bar\chi_c i \gamma_5\chi_c - \bar\chi_s i \gamma_5\chi_s)/2 -
    (\bar\chi_s\chi_c +
    \bar\chi_c\chi_s)/2\, ,  \\
    \mathcal{O}_{s}\ &\equiv\ 
    Z(\bar\chi_s i \gamma_5\chi_s - \bar\chi_c i \gamma_5\chi_c)/2 -
    (\bar\chi_s\chi_c +
    \bar\chi_c\chi_s)/2\, .  \\
  \end{split}
\end{equation}
Note that the sum of pseudo-scalar and scalar contributions appears
with the ratio of renormalisation factors $Z$, which
needs to be taken into account properly. The renormalisation factors
for eq.~(\ref{eq:physsinglet}) are the non-singlet $Z_\mathrm{P}$ and
$Z_\mathrm{S}$. While singlet and non-singlet $Z_\mathrm{P}$ are
identical, the singlet and non-singlet $Z_\mathrm{S}$ differ at two
loop order~\cite{Skouroupathis:2007jd} in perturbation
theory. Defining $\vec{\mathcal{O}}=\left(
  \mathcal{O}_{l},\mathcal{O}_{c},\mathcal{O}_{s} \right)^T$,  the
correlation function matrix is given by 
 \begin{equation}
  \label{eq:corrmatrix}
  \mathcal{C}(t) = \sum_\mathbf{x}\left< \vec{\mathcal{O}}\left(x\right)
    \otimes \vec{\mathcal{O}}\left(0\right) \right> \, . 
 \end{equation}
However, masses are independent of $Z$ as well as the choice of basis,
so, for the purpose of determining masses we can proceed as follows:
starting from the bilinears in eqs.~(\ref{eq:ll}) 
and (\ref{eq:psequ}) we change the operator basis via an appropriate
rotation matrix $\mathcal{R}$
\begin{equation}
 \label {eq:rotation1}
 \mathcal{C}_{\mathcal{R}}(t) = \left<
   \mathcal{R}\vec{\mathcal{O}}(t) \otimes
   \mathcal{R}\vec{\mathcal{O}}(0)\right> = \mathcal{R}
 \mathcal{C}(t) \mathcal{R}^T \ , 
\end{equation}
where
\begin{equation}
  \label{eq:rotatedcorrmatrix}
  \mathcal{C}_{\mathcal{R}} = 
  \begin{pmatrix}
    \eta_{\mathcal{O}_{\ell}\mathcal{O}_{\ell}} &
    \eta_{\mathcal{O}_{\ell} \mathcal{S}_h} & \eta_{\mathcal{O}_{\ell}
      \mathcal{P}_h} \\ 
    \eta_{\mathcal{S}_h \mathcal{O}_{\ell}} & \eta_{\mathcal{S}_h
      \mathcal{S}_h} & \eta_{\mathcal{S}_h \mathcal{P}_h} \\ 
    \eta_{\mathcal{P}_h \mathcal{O}_{\ell}} & \eta_{\mathcal{P}_h
      \mathcal{S}_h} & \eta_{\mathcal{P}_h \mathcal{P}_h} \\ 
  \end{pmatrix}\, ,
\end{equation}
is again a symmetric, real and positive definite correlation matrix with
\begin{equation}
  \mathcal{P}_h \equiv (\bar\chi_c i \gamma_5\chi_c - \bar\chi_s i
  \gamma_5\chi_s)/2\, ,\qquad
  \mathcal{S}_h \equiv Z^{-1} (\bar\chi_s\chi_c + \bar\chi_c\chi_s)/2
\end{equation}
and $\eta_{XY}$ denoting the corresponding correlation
function. The rotation matrix $\mathcal{R}$ is given by
\begin{equation}
 \label{eq:rotation}
 \mathcal{R} = 
  \begin{pmatrix}
   1 & 0 & 0\\
   0 & -\frac{1}{\sqrt{2}} & -\frac{1}{\sqrt{2}} \\
   0 & +\frac{1}{\sqrt{2}} & -\frac{1}{\sqrt{2}} \\
  \end{pmatrix}\,.
 \end{equation}
 Now we can drop the factor $Z^{-1}$, which appears only as a constant
scaling factor for $\mathcal{S}_h$. We denote the corresponding  
correlation matrix by $\tilde{\mathcal{C}}_{\mathcal{R}}$.
 This disentanglement of scalar and pseudo-scalar contributions
greatly reduces  the number of terms required for each element of the
correlator matrix.

Solving the generalised eigenvalue
problem~\cite{Michael:1982gb,Luscher:1990ck,Blossier:2009kd}
\begin{equation}
  \tilde{\mathcal{C}}_{\mathcal{R}}(t)\ \eta^{(n)}(t,t_0) = \lambda^{(n)}(t, t_0)\
  \tilde{\mathcal{C}}_{\mathcal{R}}(t_0)\ \eta^{(n)}(t,t_0) \,,
  \label{eq:gevp}
\end{equation}
and taking into account the periodic boundary conditions for a meson, we
can determine the effective masses by solving
\begin{equation}
  \frac{\lambda^{(n)}(t,t_0)}{\lambda^{(n)}(t+1,t_0)} = \frac{e^{-m^{(n)}
      t}+ e^{-m^{(n)}(T-t)}}
  {e^{-m^{(n)}(t+1)}+ e^{-m^{(n)}(T-(t+1))}}
\end{equation}
for $m^{(n)}$, where $n$ counts the eigenvalues. The state with the
lowest mass should correspond to the $\eta$ and the second state
to the $\eta'$ meson. Alternatively, we use a factorising fit of the
form 
\begin{equation}
  \label{eq:fit}
  \mathcal{C}_{qq'}(t) = \sum_n \frac{A_{q,n}A_{q',n}}{2 m^{(n)}}\
  \left[\exp(-m^{(n)} t) + \exp(-m^{(n)}(T-t))\right]
\end{equation}
to the correlation matrix matrix $\mathcal{C}$. For this we either
first rotate $\tilde{\mathcal{C}}_{\mathcal{R}}$ back taking the
factor $Z$ into account, see below, or we directly construct
$\mathcal{C}$ taking $Z$ into account. The amplitudes 
$A_{q,n}$ correspond to $\langle0|\bar q q|n \rangle$ with $n\equiv 
\eta,\eta',...$ and $q = \ell, s, c$.

\subsection{Flavour Content and Mixing}

One might first of all be interested in the quark flavour content of a
given state in order to compare to phenomenology. From the components
$\eta^{(n)}_{0,1,2}$ of the eigenvectors $\eta^{(n)}$ defined above,
we can reconstruct the flavour contents $c^{(n)}_{\ell,s,c}$ of the states.
Since we have changed the basis according to
eqs.~(\ref{eq:rotatedcorrmatrix}) and (\ref{eq:rotation}), we
reconstruct $c^{(n)}_{\ell,s,c}$ from
\begin{equation}
  \begin{split}
    c_\ell^{(n)} &= \frac{1}{\mathcal{N}^{(n)}} (\eta^{(n)}_0)\\
    c_c^{(n)} &= \frac{1}{\mathcal{N}^{(n)}} (- Z^{-1} \eta^{(n)}_1 +
    \eta^{(n)}_2)/\sqrt{2}\\ 
    c_s^{(n)} &= \frac{1}{\mathcal{N}^{(n)}} (- Z^{-1} \eta^{(n)}_1 -
    \eta^{(n)}_2)/\sqrt{2}\\ 
  \end{split}
\end{equation}
with normalisation
\[
\mathcal{N}^{(n)} = \sqrt{(\eta^{(n)}_0)^2 +  (Z^{-1} \eta^{(n)}_1)^2 +
  (\eta^{(n)}_2)^2}\ .
\]  
At this point the ratio $Z$ is again required. The flavour
non-singlet renormalisation factors have been 
evaluated  non-perturbatively for our situation~\cite{ETM:rimom}.
Another way to get access to these ratios of renormalisation constants
is available: we require that unphysical amplitudes (such as the
connected correlator from source $\bar{s} \Gamma s$ to sink $\bar{c}
\Gamma c$ in the physical basis), which are formally order $a$ 
contributions in the twisted mass formulation, be minimised.

For instance, for $B35$ this procedure gives $Z=0.70$, while  for
$D15$ we obtain  $Z=0.75$. For comparison the preliminary RI-MOM flavour
non-singlet value~\cite{ETM:2011aa} for $\beta=1.95$ is $0.700(8)$ and
at $\beta=2.1$ is $0.737(14)$. This close agreement indicates that we
have a reliable estimate  of $Z$.
Moreover, it turns out that the consequence of an error in this ratio on
our mixing angles is minimal, see the discussion in
section~\ref{sec:res}.

The mixing between $\eta$ and $\eta'$ is usually expressed in
terms of mixing angles in an appropriate basis. Here we use the quark
flavour basis considering only light and strange quarks. With   
 \[
|\eta_\ell\rangle\ \equiv\ \frac{1}{\sqrt{2}}(|\bar u u\rangle + |\bar
  d d\rangle)\,,\qquad\quad |\eta_s\rangle\ \equiv\ |\bar s
  s\rangle
\]
one arrives at
\[
\begin{pmatrix}
  |\eta\rangle \\
  |\eta'\rangle\\
\end{pmatrix}
=
\begin{pmatrix}
  \cos\phi & -\sin\phi \\
  \sin\phi & \cos\phi \\
\end{pmatrix}
\cdot
\begin{pmatrix}
  |\eta_\ell\rangle \\
  |\eta_s\rangle\\
\end{pmatrix}\ .
\]
For a detailed discussion including the charm quark see for
instance ref.~\cite{Feldmann:1999uf}. 

On the lattice, however, we have to work with the amplitudes $A_{q,n}$
defined above. Following refs.~\cite{Christ:2010dd,Gregory:2011sg}, we
first rotate the matrix
$\mathcal{C}_\mathcal{R}\left(t\right)$ back to the original form
$\mathcal{C}\left(t\right)$ in eq.~(\ref{eq:corrmatrix}) in the way
prescribed above (including $Z$). The amplitudes $A_{q,n}$ in
eq.~(\ref{eq:fit}) of a factorising fit to the rotated correlation
matrix are then directly related to the mixing angles
via~\cite{Feldmann:1999uf} 
\begin{equation}
  \label{eq:mixingmatrix}
  \begin{pmatrix}
    A_{\ell,\eta}  & A_{s,\eta} \\
    A_{\ell,\eta'} & A_{s,\eta'} \\  
  \end{pmatrix}
  =
  \begin{pmatrix}
    f_\ell \cos\phi_\ell & -f_s \sin\phi_s\\
    f_\ell \sin\phi_\ell & f_s \cos\phi_s\\
  \end{pmatrix}\, ,
\end{equation}
where we ignored the charm contribution. Hence, the mixing angles
$\phi_\ell$ and $\phi_s$ can be extracted from
\begin{equation}
  \label{eq:angles}
  \tan\phi_\ell = \frac{A_{\ell,\eta'}}{A_{\ell,\eta}}\,
  ,\qquad\tan\phi_s = -\frac{A_{s,\eta}}{A_{s,\eta'}}\, , 
\end{equation}
where the renormalisation constants cancel in the ratio.
Following the RBC / UKQCD and the Hadron Spectrum collaborations we
also define a common angle $\phi$ (representing the geometric mean of
$\phi_\ell$ and $\phi_s$)
\begin{equation}
  \label{eq:meanangle}
  \tan^2(\phi) = -\frac{A_{\ell\eta'}A_{s\eta}}{A_{\ell\eta}A_{s\eta'}}\,,
\end{equation}
inspired by arguments that $\phi_\ell$ and $\phi_s$ should actually
agree~\cite{Feldmann:1999uf,Leutwyler:1997yr}, see
also ref.~\cite{Gregory:2011sg}. 
We do not include the charm quark in
the discussion here, because it turns out that its contribution to
$\eta$ and $\eta'$ is negligible. 

\section{Results}
\label{sec:res}

\begin{table}[t!]
  \centering
  \begin{tabular*}{0.9\textwidth}{@{\extracolsep{\fill}}lcccccc}
    \hline\hline
    ensemble & $aM_{\mathrm{PS}}$ & $aM_\mathrm{K}$ & $aM_\eta$ & $a M_{\eta'}$  \\

    \hline\hline
    A30.32   & $0.12374(27)$ & $0.25150(29)$ & $0.286(15)$ & $0.49(6)$ \\
    A40.24   & $0.14517(39)$ & $0.25884(43)$ & $0.281(18)$ & $0.39(6)$ \\
    A40.32   & $0.14174(26)$ & $0.25666(23)$ & $0.281(11)$ & $0.49(9)$ \\
    A60.24   & $0.17340(39)$ & $0.26695(52)$ & $0.290(7)$  & $0.59(9)$ \\
    A80.24   & $0.19888(37)$ & $0.27706(61)$ & $0.302(8)$  & $0.60(8)$ \\
    A100.24  & $0.22097(40)$ & $0.28807(34)$ & $0.315(11)$ & $0.50(7)$ \\
    \hline
    A80.24s  & $0.19870(50)$ & $0.25503(33)$ & $0.270(9)$  & $0.54(9)$ \\
    A100.24s & $0.22149(39)$ & $0.26490(74)$ & $0.280(5)$  & $0.66(12)$ \\
    \hline
    B25.32   & $0.10768(30)$ & $0.21240(50)$ & $0.234(10)$ & $0.50(8)$ \\ 
    B35.32   & $0.12445(29)$ & $0.21840(28)$ & $0.237(9)$  & $0.59(9)$ \\ 
    B55.32   & $0.15503(48)$ & $0.22799(34)$ & $0.249(14)$ & $0.60(10)$ \\ 
    B75.32   & $0.18121(24)$ & $0.23753(32)$ & $0.253(13)$ & $0.44(7)$ \\ 
    B85.24   & $0.19373(58)$ & $0.24392(59)$ & $0.260(12)$ & $0.51(12)$ \\
    \hline  
    D15.48   & $0.07004(19)$ & $0.16897(85)$ & $0.201(9)$  & $0.38(8)$ \\
    D45.32sc & $0.07981(30)$ & $0.17570(84)$ & $0.192(15)$ & $0.30(4)$ \\
    \hline\hline
    \vspace*{0.1cm}
  \end{tabular*}
  \caption{Results of $a M_\eta$, $a M_{\eta'}$ for all ensembles and
    the corresponding values for the charged pion mass
    $M_{\mathrm{PS}}$ and the kaon mass $M_\mathrm{K}$. Most of the
    kaon and pion mass values have been published already in
    ref~\cite{Baron:2010bv}, however, we have recomputed the pion mass
    values with our statistics.}
  \label{tab:masses_results}
\end{table}

We have computed all contractions needed for building the correlation 
matrix of eq.~(\ref{eq:rotatedcorrmatrix}). For the connected contributions, we 
used stochastic time-slice sources (the so called
``one-end-trick''~\cite{Boucaud:2008xu}). For the disconnected
contributions, we used stochastic volume sources with complex
Gaussian noise~\cite{Boucaud:2008xu}. As discussed in
ref.~\cite{Jansen:2008wv}, one can estimate the light disconnected
contributions very efficiently using the identity
 \[
D_u^{-1} - D_d^{-1} = -2i\mu_\ell D_d^{-1}\ \gamma_5\ D_u^{-1}\ .
\]
For the heavy sector such a simple relation does not exist, but we
can use the so called hopping parameter variance reduction, which
relies on the same equality as in the mass degenerate two flavour
case (see ref.~\cite{Boucaud:2008xu} and references therein)
\[
D_h^{-1} = B - BHB + B(HB)^2 - B(HB)^3 + D_h^{-1} (HB)^4
\]
with $D_h = (1 + HB) A$, $B = 1/A$ and $H$ the two flavour hopping
matrix. The number of stochastic volume sources $N_s$ per gauge
configuration we used for both the heavy and the light sector is given
for each ensemble in table~\ref{tab:setup}. In order to check
that the stochastic noise introduced by our method is smaller than the
gauge noise we have increased $N_s$ from $24$ to $64$ for ensemble
$B25.32$. This increase in $N_s$ did not reduce the error on the
extracted masses.

\begin{table}[t!]
  \centering
  \begin{tabular*}{0.9\textwidth}{@{\extracolsep{\fill}}lcccccc}
    \hline\hline
    ensemble & $\phi_{l}^{L}$ & $\phi_{s}^{L}$ & $\phi^{L}$ &
    $\phi_{l}^{F}$ & $\phi_{s}^{F}$ & $\phi^{F}$ \\ 
    \hline\hline
    A30.32   & $57(15)$ & $49(13)$ & $56(15)$ & $36(9)$  & $53(13)$ & $46(11)$ \\
    A40.24   & $39(16)$ & $41(11)$ & $39(16)$ & $43(11)$ & $40(12)$ & $41(13)$ \\
    A40.32   & $56(18)$ & $33(11)$ & $54(18)$ & $33(11)$ & $44(11)$ & $44(11)$ \\
    A60.24   & $66(15)$ & $29(13)$ & $66(15)$ & $30(13)$ & $49(8)$  & $49(8)$  \\
    A80.24   & $73(7)$  & $20(10)$ & $73(7)$  & $24(10)$ & $48(6)$  & $50(6)$  \\
    A100.24  & $62(10)$ & $37(10)$ & $62(10)$ & $39(9)$  & $50(5)$  & $51(5)$  \\
    \hline
    A80.24s  & $59(16)$ & $36(16)$ & $59(16)$ & $41(16)$ & $48(12)$ & $50(12)$ \\
    A100.24s & $75(13)$ & $20(16)$ & $76(12)$ & $25(16)$ & $50(7)$  & $54(7)$  \\
    \hline
    B25.32   & $74(13)$ & $25(10)$ & $73(13)$ & $28(10)$ & $52(6)$  & $53(6)$  \\ 
    B35.32   & $67(14)$ & $19(7)$  & $63(15)$ & $22(8)$  & $42(6)$  & $41(7)$  \\ 
    B55.32   & $74(11)$ & $14(6)$  & $73(11)$ & $18(7)$  & $43(6)$  & $46(6)$  \\ 
    B75.32   & $50(14)$ & $52(15)$ & $50(14)$ & $52(15)$ & $51(11)$ & $51(11)$ \\ 
    B85.24   & $60(16)$ & $47(20)$ & $61(16)$ & $46(19)$ & $54(13)$ & $54(13)$ \\
    \hline  
    D15.48   & $59(14)$ & $25(9)$  & $57(14)$ & $29(9)$  & $41(10)$ & $42(10)$ \\
    D45.32sc & $60(19)$ & $57(15)$ & $59(19)$ & $58(15)$ & $59(15)$ & $58(15)$ \\
    \hline\hline
    \vspace*{0.1cm}
  \end{tabular*}
  \caption{Results for the mixing angles in eqs.~(\ref{eq:angles}) and
    (\ref{eq:meanangle}) from a $4\times4$-correlation function matrix
    using local (L) and fuzzed (F) operators for all ensembles} 
  \label{tab:angles_results}
\end{table}

Also, we use both local and fuzzed operators to enlarge our correlation
matrix by a factor two. In addition to the interpolating operator
quoted in eqs.~(\ref{eq:ll}) and (\ref{eq:physsinglet}),  one could also
consider the $\gamma$-matrix combination $i \gamma_0\gamma_5$, which would
increase the correlation matrix by another factor of two. However, the
corresponding correlation functions turn out to be too noisy to give
any further improvement at this stage. The values for the number of 
gauge configurations $N_\mathrm{conf}$ investigated per ensemble are
summarised in table~\ref{tab:setup}. 

Errors are always computed using a bootstrap procedure
with $1000$ bootstrap samples. To account for autocorrelation we block
the data in blocks of length $N_b$. The values of $N_b$, see
table~\ref{tab:setup}, have been chosen such that the blocks are 
statistically independent. As autocorrelation appears to be
significant -- in particular for $M_{\eta'}$ -- we have computed the
integrated autocorrelation time $\tau_\mathrm{int}$ in units of HMC
trajectories of length $1$ using the
$\Gamma$-method~\cite{Wolff:2003sm} for the elements of the
correlation matrix $\mathcal{C_R}$ at fixed time $t/a=3$ for 
several ensembles. Most affected by autocorrelation are the matrix
elements with light quark content. All other elements of
$\mathcal{C}_\mathcal{R}$ are only mildly affected. For instance, for
ensemble $D15.48$ the matrix element with only light quark content has
integrated autocorrelation time of $\tau_\mathrm{int} = 9(2)$, while
the elements without light quark content have at most
$\tau_\mathrm{int} = 1.3(2)$. Note that our normalisation is such that
$\tau_\mathrm{int}=0.5$ corresponds to no
autocorrelation. Alternatively to the $\Gamma$-method we have varied 
the blocklength of the bootstrap method and found that for a
blocklength of $N_b=10$ or larger the error for all matrix elements
for ensemble $D15.48$ stays constant within error. From the latter
method we obtain $\tau_\mathrm{int} = 7(2)$ for the matrix
elements with light quark content consistent with the result from the
$\Gamma$-method. $N_b=10$ corresponds for $D15.48$ to $20$ HMC
trajectories of length $1$ (see ref.~\cite{Urbach:2005ji} for a
description of the HMC algorithm used).

The autocorrelation depends on the lattice spacing. With increasing
value of the lattice spacing the autocorrelation becomes less. For
instance, the light-only matrix element has $\tau_\mathrm{int}=6(1)$
for $B25.32$ and $\tau_\mathrm{int}=4(1)$ for $A30.32$, compared to
the aforementioned $\tau_\mathrm{int}=9(2)$ for $D15.48$. The quark
mass dependence is not significant.

The details of our GEVP and fitting procedures to extract $\eta$ and
$\eta'$ masses are explained in appendix~\ref{sec:app}, together with
fit ranges and $\chi^2$ values. For the masses we used only the
blocked bootstrap method to analyse autocorrelation and found that the
$\eta'$ state (in agreement to what we found  in the $N_f=2$
case~\cite{Jansen:2008wv}) shows significant autocorrelation, whereas
the $\eta$ is less affected. Again, for ensemble $D15.48$ a blocksize
of $N_b=10$ seems to yield statistically independent blocks. We
emphasise that with our values of $N_\mathrm{conf}$ we can hardly use
$N_b>20$, because the number of blocks becomes too small. Therefore,
we cannot exclude that there is autocorrelation on longer scales.

\subsection{Extraction of Masses}

\begin{figure}[t]
  \centering
  \subfigure[]{\includegraphics[width=.48\linewidth]
    {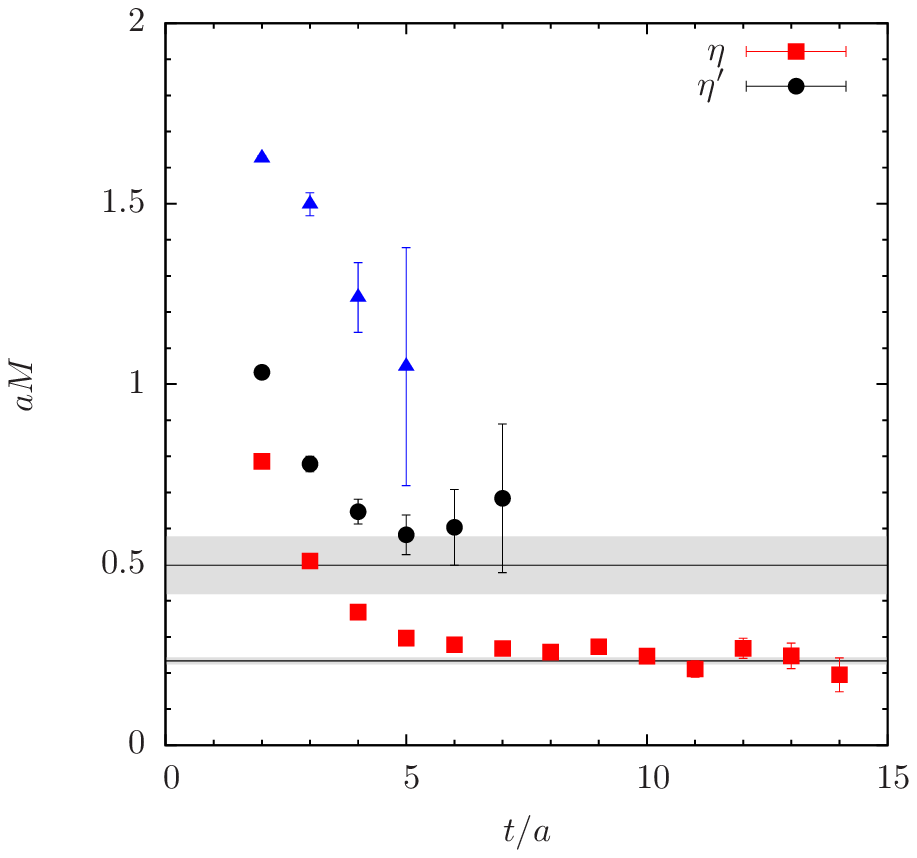}}\quad
  \subfigure[]{\includegraphics[width=.48\linewidth]
    {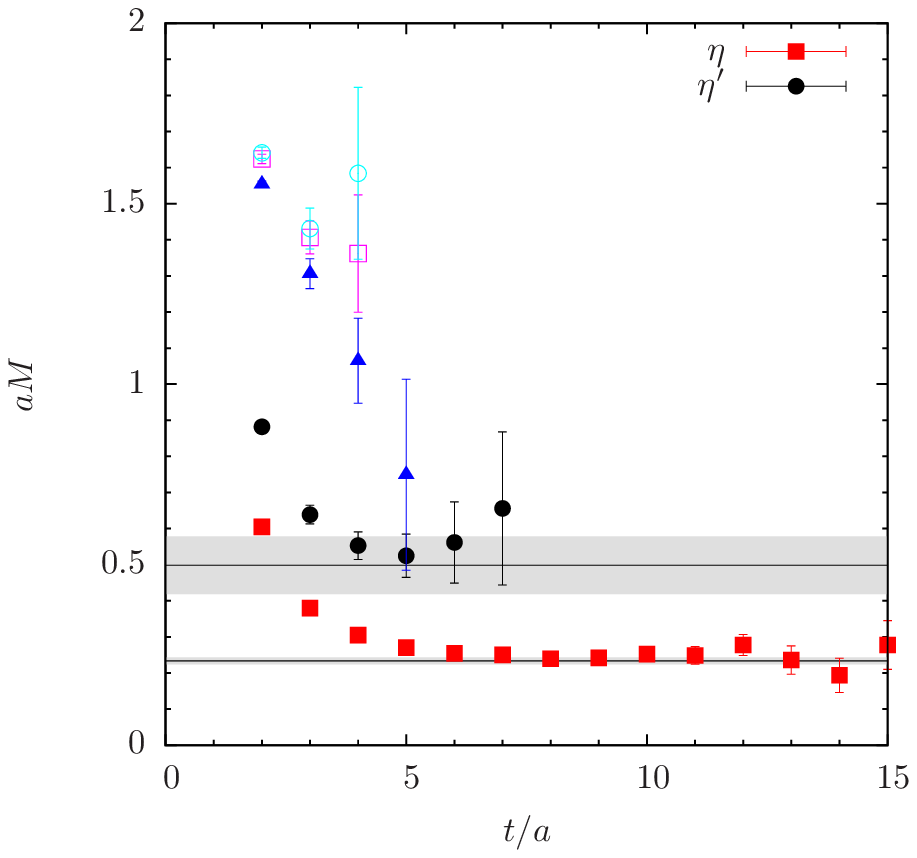}}\quad
  \caption{(a) Effective masses in lattice units determined from solving
    the GEVP for a $3\times3$ matrix with $t_0/a=1$ for ensemble
    B25.32. (b) the same, but for a $6\times6$ matrix. For comparison
    also the fit results (see text) for the two lowest states are
    shown (cf. table~\ref{tab:masses_results}).} 
  \label{fig:effmass}
\end{figure}

In the left panel of figure~\ref{fig:effmass}, we show the effective
masses determined from solving the GEVP for ensemble
$B25.32$ from a $3\times3$ matrix with local operators only. For the
purpose of this plot, we kept 
$t_0/a=1$ fixed. One observes that the ground state is very well
determined and it can be extracted from a plateau fit. The second
state, i.e. the  $\eta'$, is much more noisy and a mass determination
is questionable, at least from a $3\times3$ matrix. Enlarging the
matrix size  significantly reduces the contributions of excited states
to the lowest states and, due to smaller statistical errors at smaller
$t$ values, a determination becomes possible. This can be seen for a
$6\times6$ (including fuzzing) matrix in the right panel of
figure~\ref{fig:effmass}. Our final results and errors for the masses
are indicated in figure~\ref{fig:effmass} by the horizontal bands. We
do not determine them from a constant fit to the plateau, but by a
three state $\cosh$ fit to the eigenvalues, possibly with larger $t_0$
values, see appendix~\ref{sec:app}. As shown, the 
procedure gives very good agreement with a plateau fit for the $\eta$,
but slightly lower values for the $\eta'$. This indicates non-negligible
systematic uncertainties in the extraction of the $\eta'$ mass value.

The third state appears to be in the region where one would expect the
$\eta_c$ mass value, however, the signal is lost at $t/a=5$,
which makes a reliable determination unfeasible. Note that the two
plots in figure~\ref{fig:effmass} are rather independent of the
particular ensemble chosen.

\begin{figure}[t]
  \centering
  \subfigure[]{\includegraphics[width=.48\linewidth]
  {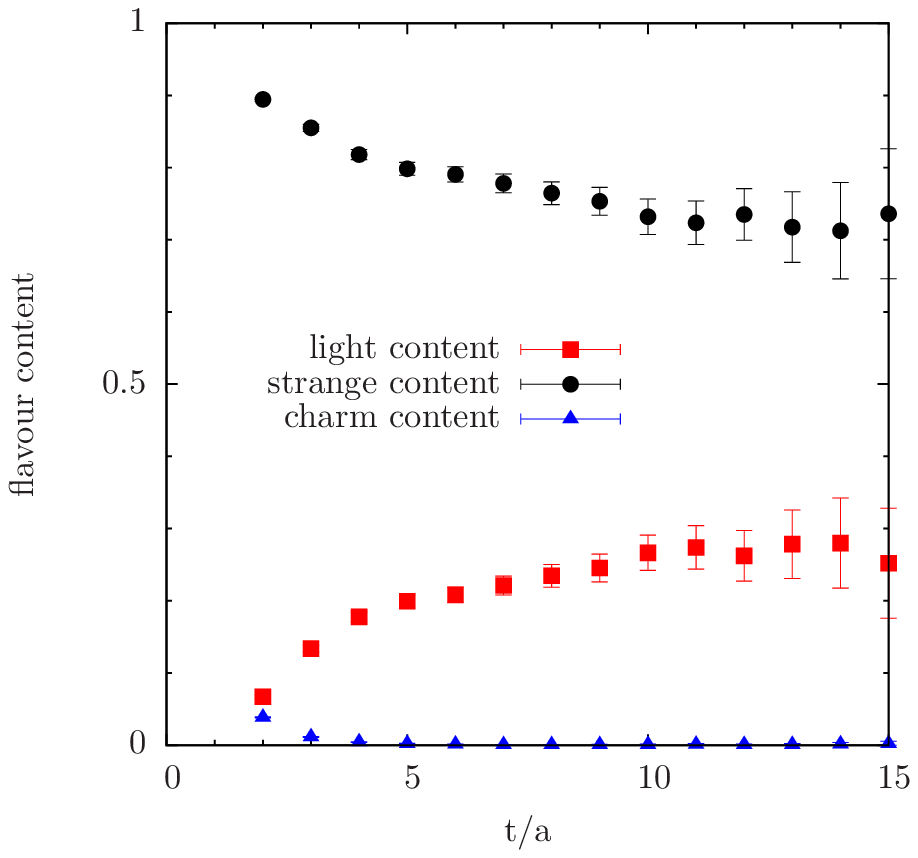}}\quad
  \subfigure[]{\includegraphics[width=.48\linewidth]
  {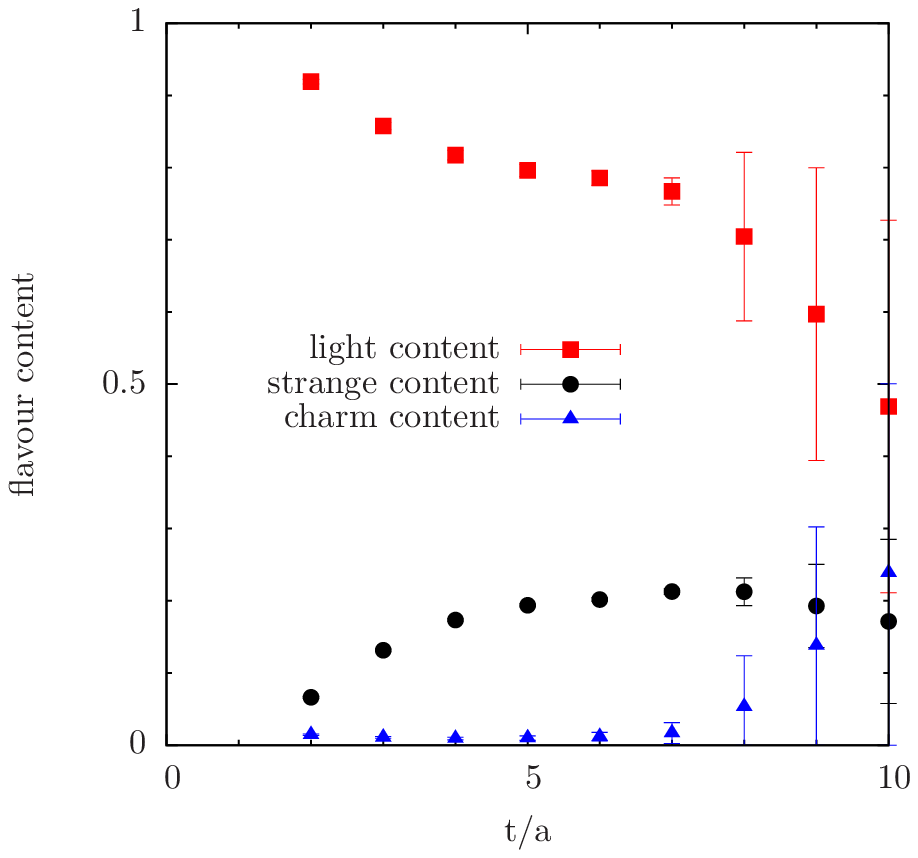}}
  \caption{(a) Squared flavour content of $\eta$ for B25.32 from
    $3\times3$-matrix using local operators only. (b) Squared flavour
    content of $\eta'$ for B25.32.}  
  \label{fig:flavourcontent}
\end{figure}

To gain further confidence in our identification of the $\eta$ and
$\eta'$ states, we also determine the flavour content of the two
states as explained above. As an example we show in the left panel of
figure~\ref{fig:flavourcontent} the flavour content of the $\eta$ for
ensemble $B25.32$. It becomes evident that -- as one would expect
from phenomenology -- the $\eta$ has a dominant strange quark content,
while the  $\eta'$, shown in the right panel of
figure~\ref{fig:flavourcontent} is dominated by light quarks. Note
that we do not include a gluonic operator in our analysis, so we
only discuss the relative quark content. For
both, the charm contribution is compatible with zero.

\begin{figure}[t]
  \centering
  \includegraphics[width=.7\linewidth]%
  {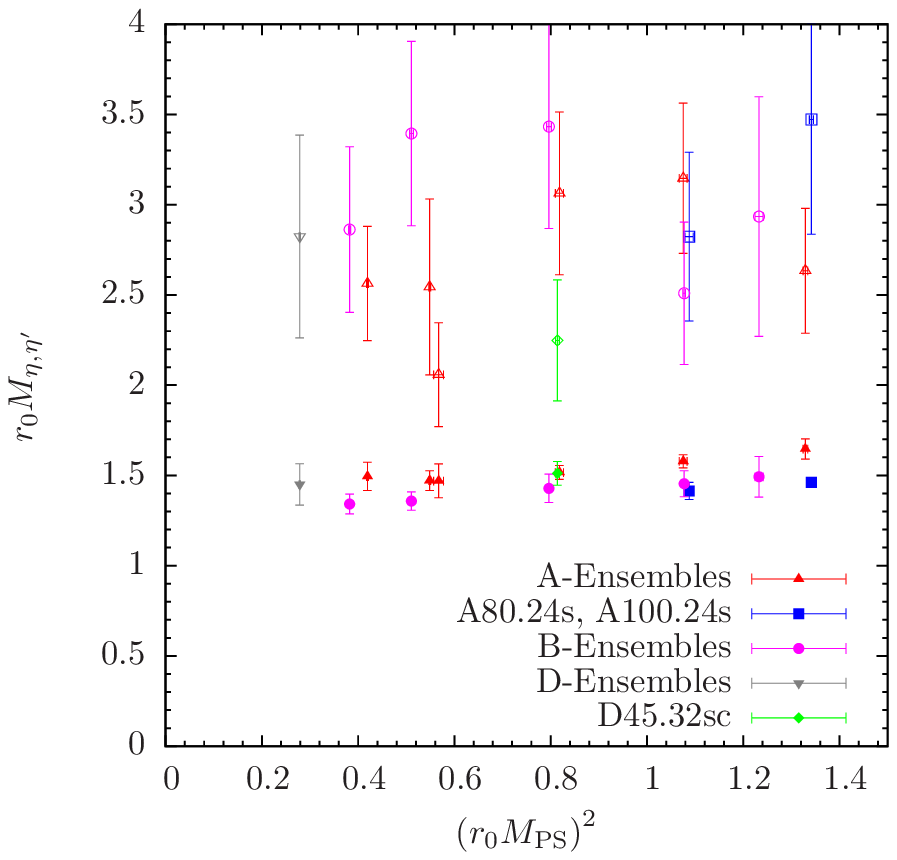}
  \caption{(a) $\eta$ (filled symbols) and $\eta'$ (open symbols)
    masses in units of chirally extrapolated $r_0$ (listed in
    table~\ref{tab:r0values}) as a function of $(r_0 
    M_\mathrm{PS})^2$.}
  \label{fig:masssummary}
\end{figure}

In figure~\ref{fig:masssummary} we show the masses
of the $\eta$ (filled symbols) and $\eta'$ (open symbols) mesons for
the various ensembles we used as a function of the 
squared pion mass, everything in units of $r_0$. All the masses have
been determined from solving the GEVP for a $6\times6$ matrix, the
details are described in appendix~\ref{sec:app}. The results have been
independently cross-checked using a factorising fit, and corroborated.
We have collected all the values for $aM_\eta$ and
$aM_{\eta'}$ together with kaon and pion mass values in
table~\ref{tab:masses_results} with statistical errors only. Note that
the values for $aM_\mathrm{K}$ and $aM_\mathrm{PS}$ are published for
most of the ensembles already in ref.~\cite{Baron:2010bv}. But we have
recomputed the pion mass values with our statistics, and added
ensembles as compared to ref.~\cite{Baron:2010bv}. The methods for
computing the kaon mass in $N_f=2+1+1$ Wilson twisted mass lattice QCD
are described in ref.~\cite{Baron:2010th}.
  
It is clear from the figure that the $\eta$ meson
mass can be extracted with high precision, while the $\eta'$ meson
mass is more noisy. The former can be understood, because in the SU(3)
symmetric limit the $\eta$ meson is a flavour octet with all
disconnected contributions vanishing, while the $\eta'$ is the flavour
singlet with non-vanishing disconnected contributions\footnote{We
  thank Martin Savage for a useful discussion on this point.}. 

\begin{figure}[t]
  \centering
  \subfigure[]{\includegraphics[width=.48\linewidth]%
    {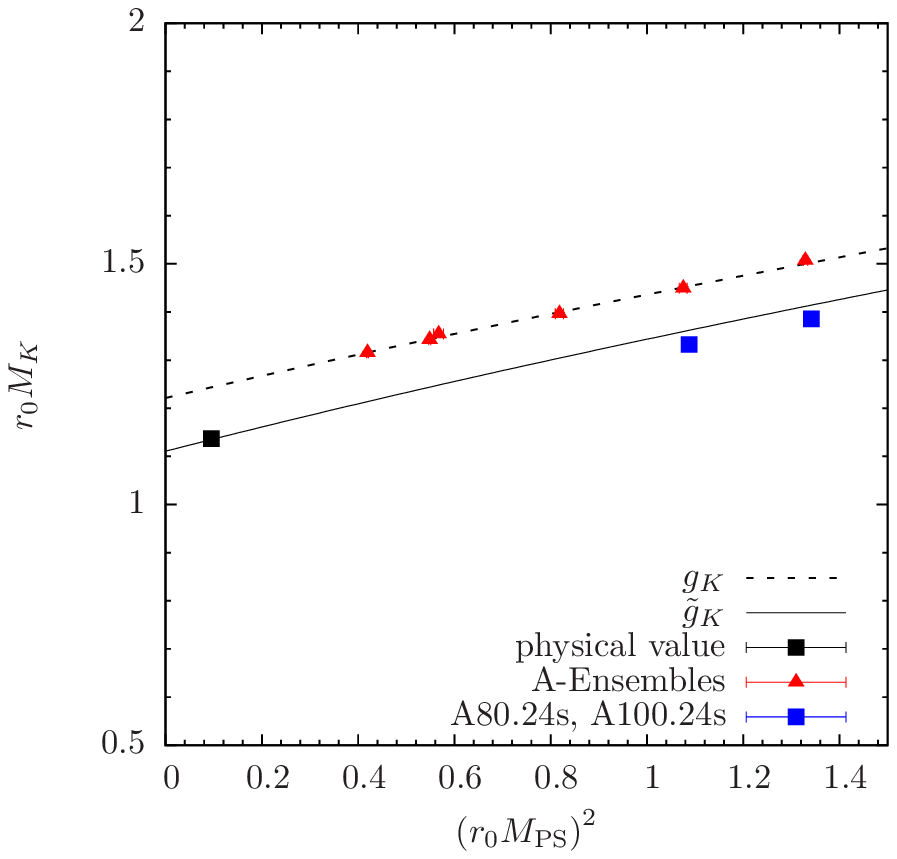}}
  \subfigure[]{\includegraphics[width=.48\linewidth]%
    {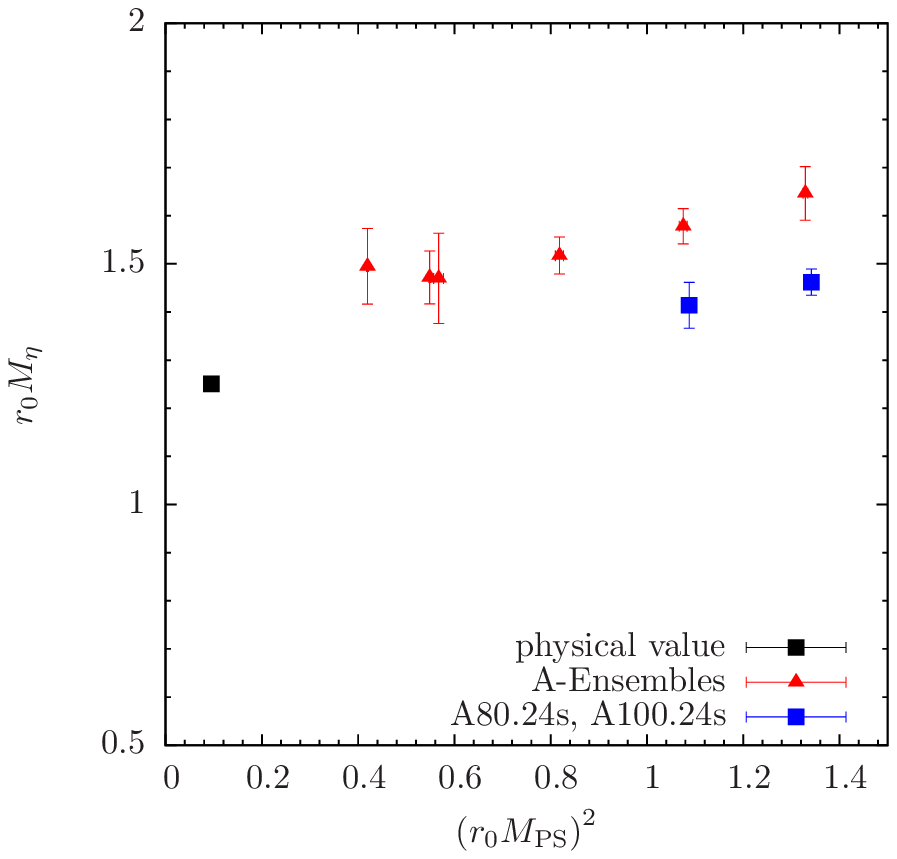}}  
  \caption{(a) The kaon mass in 
    units of $r_0$ and (b) $r_0 M_\eta$ as a function of $(r_0
    M_\mathrm{PS})^2$ also for all $A$ ensembles. The dotted and solid
    curves in (a) represent the fitted $g_K$ in eq.~(\ref{eq:g_K}) and
    the shifted $\tilde{g}_K$ (see text), respectively.}
  \label{fig:metaA}
\end{figure}

The results displayed in figure~\ref{fig:masssummary} have been
obtained using the bare values of $a\mu_\sigma$ and $a\mu_\delta$ as
used for the production of the ensembles. Those values, however, did
not lead to the physical values of, e.g., the kaon and D-meson
masses~\cite{Baron:2010bv,Baron:2010th}. We show the kaon mass as a
function of the squared pion mass, in the left
panel of  figure~\ref{fig:metaA}, for all $A$ ensembles. It is clearly
evident that the re-tuned ensembles ($A80.24s$ and $A100.24s$) have values
for the kaon mass closer to the physical one (see also
e.g. ref.~\cite{Baron:2010bv,Baron:2011sf}). In the 
right panel of figure~\ref{fig:metaA} we show $r_0M_\eta$ also for all
$A$ ensembles (including $A80.24s$ and $A100.24s$), and the same pattern as
for the kaon mass is observed. 
Furthermore, the physical strange and charm quark mass values differ
among the $A$, $B$ and $D$ ensembles. Hence,
figure~\ref{fig:masssummary} is not yet conclusive 
with regards to the size of lattice artifacts and the extrapolation to
the physical point. What we can recognise is that the light quark mass
dependence in the $\eta$ appears to be rather weak. 

While the results for the $\eta$ mass in figure~\ref{fig:masssummary}
show a consistent picture over all lattice spacings and light quark
mass values -- keeping the differences in the
strange and charm quark masses in mind -- the $\eta'$ mass shows large
fluctuations and no consistent picture. We attribute this to two
observations: firstly, due to the large noise in the $\eta'$ state,
the extracted masses are for most of the ensembles only an upper bound
for $M_{\eta'}$, because a plateau is hardly visible. Therefore, we
think the values of the $\eta'$ mass very likely have a 
non-negligible systematic uncertainty stemming from the fact that the
signal for the $\eta'$ is lost in noise at rather small $t$-values.
We are currently investigating a solver for the GEVP using a singular
value decomposition, which is better suited to deal with the
noise. First results are encouraging.

Secondly, the strange (and charm) quark mass values are
different among the three values of the lattice spacing. The 
strange quark mass value has a strong influence on the $\eta'$ meson mass,
as we learn from $A80.24$, $A80s$, $A100.24$ and $A100.24s$, see
figure~\ref{fig:masssummary}. 

Finally, as some of our ensembles have a value of $M_\mathrm{PS}\cdot
L<3.5$ it is interesting to estimate finite size corrections to
$M_\eta$ and $M_{\eta'}$. We have two ensembles which differ only
in $L/a$, namely $A40.24$ and $A40.32$ with $M_\mathrm{PS}\cdot L=3.5$
and $M_\mathrm{PS}\cdot L = 4.5$, respectively. For these two ensembles
both masses agree within errors. In particular, $M_\eta$ agrees
precisely, as can be seen from table~\ref{tab:masses_results}. On the same
ensembles we measure a finite-size effect for the kaon of below
1\%. Therefore, we conclude that at our current level of precision
finite size corrections to $M_\eta$ and $M_{\eta'}$ are not
significant. Of course, for a definite conclusion more ensembles with
different $L/a$-values are needed. 

\subsection[Scaling Artifacts and Strange Quark Mass Dependence of
  the eta mass]
{Scaling Artifacts and Strange Quark Mass Dependence of $M_\eta$}

For $M_\eta$ the statistical uncertainty is sufficiently small to
allow for a
meaningful scaling test. For this we need to compare $M_\eta$ at
the three different values of the lattice spacing 
for fixed values of for instance $r_0 M_\mathrm{K}$, $r_0
M_\mathrm{D}$, 
$r_0M_\mathrm{PS}$ and the physical volume. From volume and the
charm quark mass value we expect only
little influence given our uncertainties and hence, we are going to
disregard effects from slightly different physical volumes at the
different $\beta$-values and the differences in the charm quark
mass in the following. 

As we do not have simulations at the three values of the lattice
spacing with matched values of $r_0 M_\mathrm{K}$, we have to perform
an interpolation in $M_\mathrm{K}$. For this procedure we have to rely
on two pairs of ensembles, namely ($A80.24$, $A80.24s$) and
($A100.24$, $A100.24s$). The two ensembles within a pair differ in the
values of $a\mu_\sigma$ and $a\mu_\delta$, whereas $a\mu_\ell$ is
identical. We can use these
ensembles to estimate the derivative $D_\eta$ of $M_\eta^2$ with
respect to $M_\mathrm{K}^2$ and use this estimate to correct for the
mismatch in $r_0 M_\mathrm{K}$. By using this estimate of the
derivative for all ensembles -- not only $A80.24$ and $A100.24$ -- we neglect the
dependence of $D_\eta$ on the lattice spacing and the light and
charm quark masses.

\begin{figure}[t]
  \centering
  \subfigure[]{\includegraphics[width=.48\linewidth]{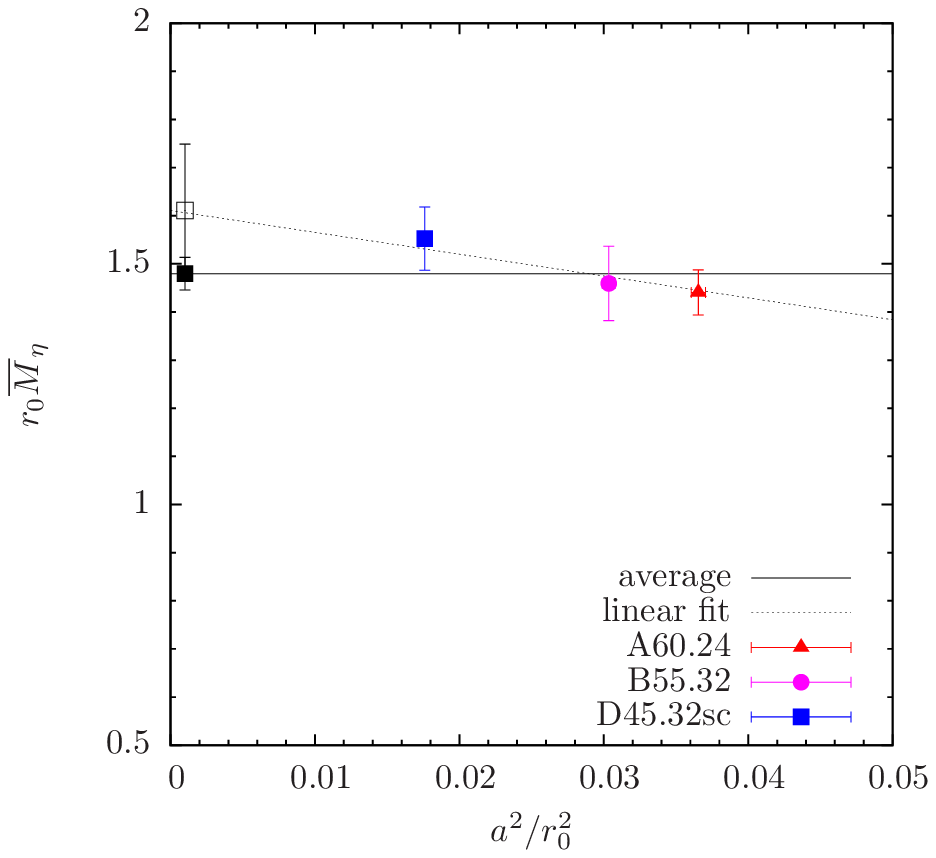}}
  \quad
  \subfigure[]{\includegraphics[width=.48\linewidth]{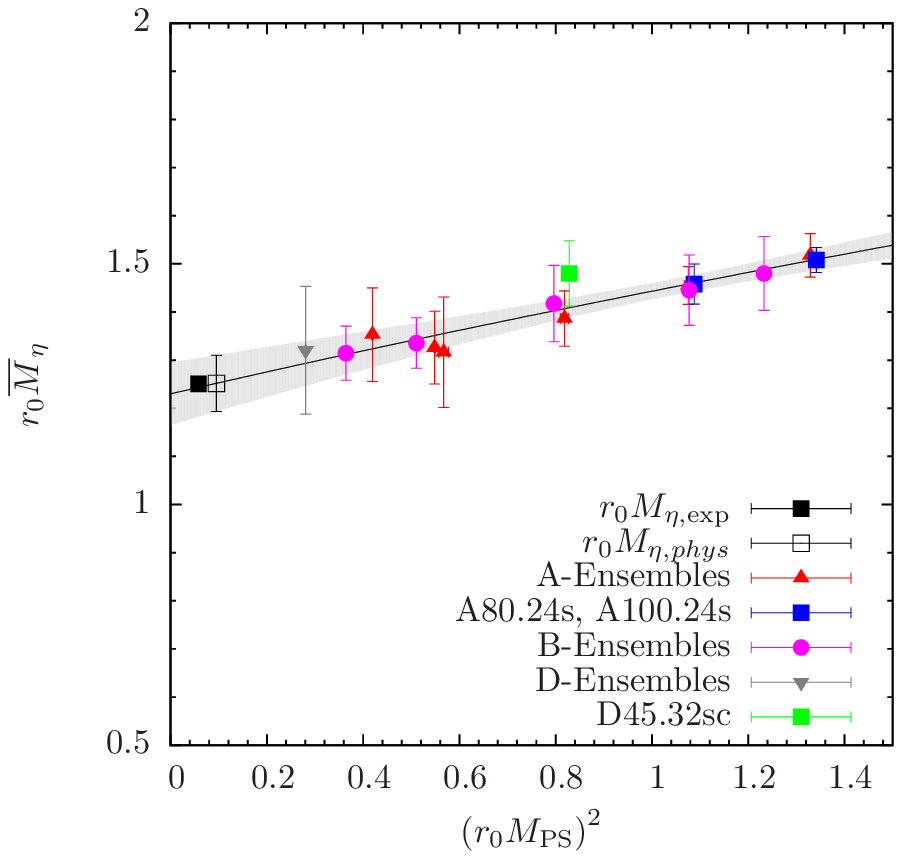}}
  \caption{ (a) $r_0\overline{M}_\eta$ as a function of $(a/r_0)^2$
    for the ensembles $A60.24$, $B55.32$ and $D45.32sc$. (b) Values of
    $r_0 \overline{M}_\eta$ as a function of $(r_0 M_\mathrm{PS})^2$
    as explained in the text. The continuum extrapolated values $r_0
    M_{\eta}^{a\rightarrow0}$ in (a) and the experimental value
    $r_{0}M_{\eta,\mathrm{exp}}$, obtained with $r_0=0.45(2)\
    \mathrm{fm}$ in (b) are horizontally displaced for legibility.}
  \label{fig:scaling}
\end{figure}

In more detail, we treat the masses of the $\eta$-meson and the kaon
like in chiral perturbation theory as functions $M^2 =
M^2[M_\mathrm{PS}^2, M_\mathrm{K}^2]$ and define the dimensionless
quantity 
\begin{equation}
  \label{eq:Deta}
  D_\eta(\mu_\ell, \mu_\sigma, \mu_\delta, \beta)\ \equiv\
  \left[\frac{d (a M_\eta)^2}{d (a M_\mathrm{K})^2}\right]\ .
\end{equation}
Next we make the approximation that
$D_\eta$ is independent of the quark mass values $\mu_\ell,
\mu_\sigma, \mu_\delta$ and $\beta$. Its value is actually equal
within errors when estimated from $A80.24$ and $A80.24s$ or from
$A100.24$ and $A100.24s$ and on average we obtain $D_\eta =
1.60(18)$. 

Now we use this value of $D_\eta$ to correct the three ensembles $A60$,
$B55$ and $D45$ -- which have approximately equal values of $r_0
M_\mathrm{PS}\approx0.9$ -- to a common value of $r_0
M_\mathrm{K}\approx1.34$ using
\[
(r_0 \overline{M}_\eta)^2 = (r_0 M_\eta)^2 +
  D_\eta\cdot\Delta_\mathrm{K}\, ,
\]
where $\Delta_\mathrm{K}$ is the difference in the squared kaon mass
values (in units of $r_0$).
We plot the resulting $r_0 \overline{M}_\eta$ values for the three ensembles
$A60$, $B55$ and $D45$ as a function of $(a/r_0)^2$ in the left panel
of figure~\ref{fig:scaling}. The data are compatible with a constant 
continuum extrapolation $r_0 M_{\eta,\mathrm{const}}^{a\rightarrow0} =
1.480(34)$, which we indicate by the horizontal line. We 
can also attempt a linear extrapolation, which is also shown in the
figure, leading to $r_0 M_{\eta,\mathrm{lin}}^{a\rightarrow0} =
1.61(14)$. The difference in between the two extrapolated values 
\begin{equation}
r_0 \Delta M_\eta^{a\rightarrow0} = 0.13(13)
\label{eq:deltascaling}
\end{equation}
gives us an estimate on the systematic uncertainty to be expected from the
continuum extrapolation. We cannot repeat this analysis for more
values of $r_0M_\mathrm{PS}$, as we have currently only two $D$
ensembles analysed. We will therefore quote a $8\%$ relative error
from $\Delta
M_\eta^{a\rightarrow0}/M_{\eta,\mathrm{const}}^{a\rightarrow0}$ for
our mass estimates, which we believe is a conservative figure.

Nevertheless, in order to obtain a more complete picture, we now attempt to
correct all our ensembles for the slightly miss-tuned value of
$M_\mathrm{K}$. For this we adopt the following procedure: first we
perform a linear fit
\begin{equation}
  \label{eq:g_K}
  g_\mathrm{K}[(r_0M_\mathrm{PS})^2,a,b] = a + b\cdot(r_0M_\mathrm{PS})^2
\end{equation}
to the values of $(r_0 M_K)^2$ for the $A$-ensembles (without
$A80.24s$ and $A100.24s$). We obtain $a=1.492(6)$ and $b=0.571(8)$. 
This curve is then shifted to a value $\tilde a=1.238(6)$
such that $(r_0M_\mathrm{K}^\mathrm{exp})^2 =
g_\mathrm{K}[(r_0M_\pi)^2,\tilde a,b]$. We denote this new function by
$\tilde{g}_\mathrm{K}=g_\mathrm{K}[(r_0M_\mathrm{PS})^2, \tilde
a,b]$. The two curves corresponding to $g_K$ and $\tilde{g}_K$ are
shown in the left panel of figure~\ref{fig:metaA}. As it turns out,
the kaon masses for the two tuned ensembles $A80.24s$ and $A100.24s$
are already very close to $\tilde{g}_K$. 
Now we can correct the remaining $A$ ensembles as well as the $B$ and
$D$ ensembles to the same line of $\tilde{g}$ by computing
the difference of the squared kaon mass values to $\tilde{g}$
\[
\delta_\mathrm{K}[(r_0M_\mathrm{PS})^2] = (r_0 M_\mathrm{K})^2
[(r_0M_\mathrm{PS})^2] -  
\tilde{g}_\mathrm{K}[(r_0M_\mathrm{PS})^2]\,,
\]
which we can use to correct the measured
$M_\eta^2[(r_0M_\mathrm{PS})^2]$ corresponding to  
\begin{equation}
  \label{eq:correction}
  (r_0\overline{M}_\eta)^2[(r_0M_\mathrm{PS})^2] = (r_0
  M_\eta)^2[(r_0M_\mathrm{PS})^2] +
  D_\eta\cdot\delta_\mathrm{K}[(r_0M_\mathrm{PS})^2]\, .
\end{equation}
The result of this procedure is shown in the right panel of
figure~\ref{fig:scaling}: we show values of $r_0 \overline{M}_\eta$  
for all our ensembles as a function of $(r_0 M_\mathrm{PS})^2$. It is
evident that all the data fall on a single curve within statistical
uncertainties. 
Figure~\ref{fig:scaling} confirms that $M_\eta$ is not affected
by large cut-off effects. Note again that we ignored the
$\mu_\ell$, $\mu_\sigma$, $\mu_\delta$ and $\beta$ dependence of
$D_\eta$ and that we cannot fully estimate the systematics stemming
from this approximation.

\subsection{Extrapolation to the Physical Point}

As $\tilde{g}[(r_0M_\pi)^2]=(r_0M_\mathrm{K}^\mathrm{exp})^2$ --
and the strange quark mass was fixed to its physical value using
$M_\mathrm{K}^\mathrm{exp}$ -- we can next attempt a linear fit to all
corrected (by the procedure discussed above) data points for $(r_0
\overline{M}_\eta)^2[(r_0 M_\mathrm{PS})^2]$. Using $r_0=0.45(2)$ as
discussed in
section~\ref{sec:intro}, the fit yields $r_0 M_{\eta}\left[r_0^2
  M_\pi^2\right] = 1.252(58)_\mathrm{stat}(100)_\mathrm{sys}$ and in
physical units 
\[
M_\eta(M_\pi) = 549(33)_\mathrm{stat}(44)_\mathrm{sys} \ \mathrm{MeV}  \ ,
\]
where the experimental mass-value of the neutral pion $M_{\pi^0} =
135\ \mathrm{MeV}$ has been used for $M_\pi$. 
In the $SU(2)$ chiral limit we obtain $r_0M_\eta^0 =
1.230(65)_\mathrm{stat}(98)_\mathrm{sys}$ or  
\[
M_\eta^0 = 539(35)_\mathrm{stat}(43)_\mathrm{sys} \ \mathrm{MeV} \ .
\]
For estimating the systematic error we used the ratio $\Delta
M_\eta^{a\rightarrow0}/M_{\eta,\mathrm{const}}^{a\rightarrow0}$ as
discussed above. Note that the error on $r_0=0.45(2)$ significantly
contributes to the statistical errors for the results in physical
units. 

\begin{figure}[t]
  \centering
  \subfigure[]{\includegraphics[width=0.48\linewidth]%
    {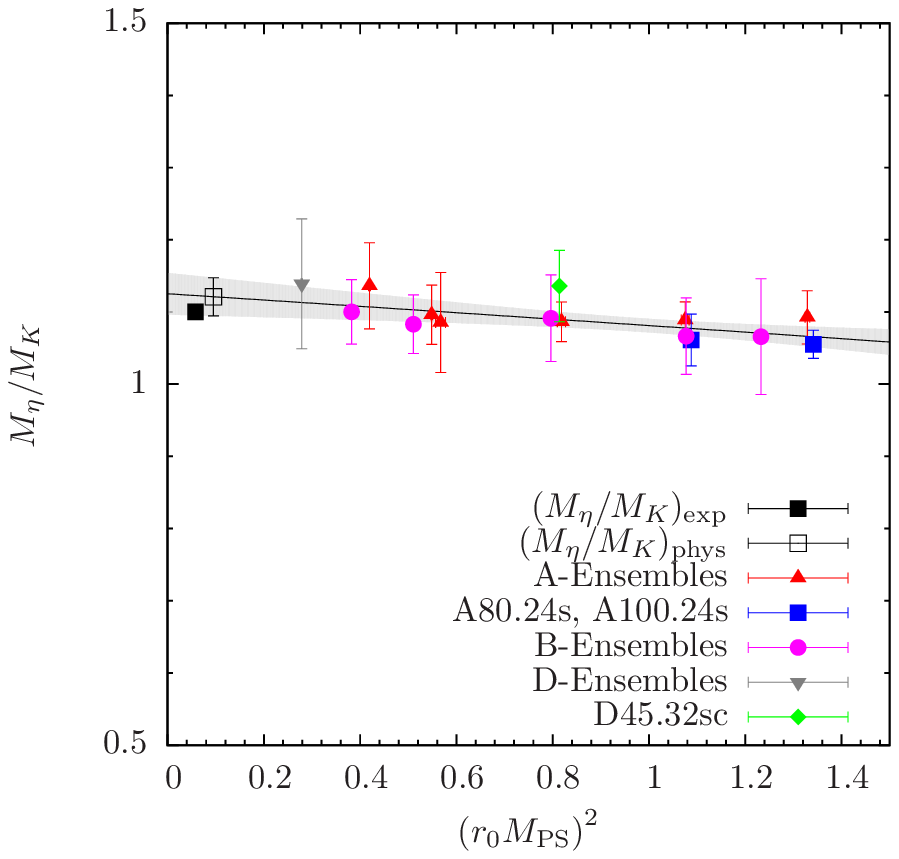}}\quad
  \subfigure[]{\includegraphics[width=0.48\linewidth]%
    {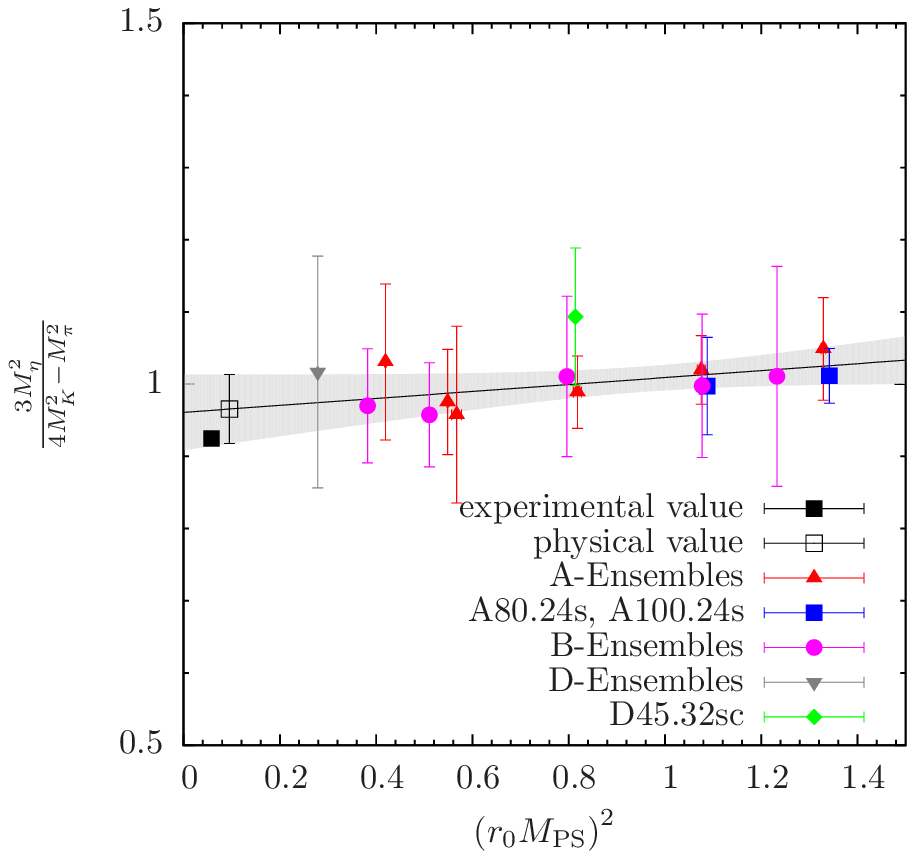}}\quad
  \caption{(a) $M_\eta/M_\mathrm{K}$ as a function of $(r_0
    M_\mathrm{PS})^2$ for all available ensembles. (b) GMO ratio (see
    text) as a function of $(r_0M_\mathrm{PS})^2$ for all available
    ensembles. Experimental values are horizontally displaced for legibility.}
  \label{fig:Mratio}
\end{figure}

As the procedure presented above relies on the assumption that
$D_\eta$ is independent on $\mu_\ell$, $\mu_\sigma$, $\mu_\delta$ and
$\beta$, it would be desirable to have a cross-check. 
Inspecting figure~\ref{fig:metaA} once more, one observes that
$M_\eta$ and $M_\mathrm{K}$ appear to have a similar strange quark mass
dependence. This motivates to search for an appropriate ratio of
quantities in which the strange quark mass dependence cancels
approximately. One option is to study the ratio $M_\eta/M_\mathrm{K}$,
which is shown in the left panel of figure~\ref{fig:Mratio} as a
function of $(r_0M_\mathrm{PS})^2$ for all available
ensembles. Strikingly, within errors all data points fall on the same
curve, and in particular the points for $A80.24s$ and $A80.24s$ and $A100.24$
and $A100.24s$ agree within errors, respectively. This confirms in
particular that most of the strange quark mass dependence cancels in
the ratio. Extrapolating all the data for $(M_\eta/M_\mathrm{K})^2$
linear in $(r_0M_\mathrm{PS})^2$ to the physical pion mass point we
obtain
\begin{equation}
 \left(M_\eta/M_\mathrm{K}\right)_{M_\pi}= 1.121(26) \ ,
\end{equation}
which is in good agreement with the experimental value
$\left(M_\eta/M_\mathrm{K}\right)_\mathrm{exp} = 1.100$. Using the experimental
value of $M_\mathrm{K^0}=498\ \mathrm{MeV}$ we obtain
\[
M_\eta = 558(13)_\mathrm{stat}(45)_\mathrm{sys} \ \mathrm{MeV}\,,
\]
where the first error is statistical and the second systematical
as estimated from the scaling violations discussed above. The latter
represents a rather conservative estimate since some of the scale
dependence might cancel in the ratio $M_\eta/M_\mathrm{K}$. Note that
in this procedure the scale $r_0=0.45(2)\ \mathrm{fm}$ enters only for
determining the physical pion mass point. As the slope of the extrapolation is
rather small, the statistical uncertainty in $M_\eta$ is smaller than
for the direct extrapolation of $(r_0M_\eta)^2$.

Alternatively, motivated from chiral perturbation theory, we also study
the GMO relation
\begin{equation}
  \label{eq:GMOmasses}
  3M_\eta^2 = 4M_\mathrm{K}^2 - M_\pi^2\,.
\end{equation}
It is valid in the SU$(3)$ symmetric case, but violated only by a few
percent with physical values of the corresponding meson
masses. Therefore, the strange quark mass dependence of
$3M_\eta^2/(4M_\mathrm{K}^2 - M_\pi^2)$ should be weak. We
show the dimensionless ratio $3M_\eta^2/(4M_\mathrm{K}^2 - M_\pi^2)$ as a
function of $(r_0 M_\mathrm{PS})^2$ in the right panel of
figure~\ref{fig:Mratio}. The first interesting remark on this figure
is that again the data points for the ratio from $A80.24$ and $A80.24s$
($A100.24$ and $A100.24s$) -- which differ in the bare strange quark mass --
agree within errors, confirming that a large part of the 
strange quark mass dependence cancels in the ratio (like for
$M_\eta/M_\mathrm{K}$).
Again, all the data points fall onto one single
line within errors, independent of the value of the lattice
spacing, the strange and the charm quark mass. If we fit a linear
function in $(r_0M_\mathrm{PS})^2$ to the data we obtain
\begin{equation}
 \left(3M_\eta^2/(4M_\mathrm{K}^2 - M_\pi^2)\right)_{M_\pi} = 0.966(48)
\end{equation}
at the physical pion mass, which is in agreement with
experiment, $(3M_\eta^2/(4M_\mathrm{K}^2 - M_\pi^2))^\mathrm{exp} = 0.925$.
Using the experimental values of $M_{\pi^0}$ and $M_{\mathrm{K}^0}$ we
now obtain 
\[
M_\eta = 559(14)_\mathrm{stat}(45)_\mathrm{sys}\ \mathrm{MeV}\,,
\]
where the first error is statistical and the second systematical
estimated from the scaling violations discussed above.

We remark that our value for $D_\eta=1.60(18)$ is in rather good
agreement to the value $4/3$ one would obtain naively from
eq.~(\ref{eq:GMOmasses}).

In the results we quoted for the physical value of $M_\eta$ we
specified a systematic uncertainty of 8\% stemming from the continuum
extrapolation. As we have some ensembles with small values of
$M_\mathrm{PS}\cdot L$, one might wonder how finite size corrections
to $M_\mathrm{K}$ and $M_\mathrm{PS}$ influence our
extrapolations. This influence is smaller than our statistical
uncertainty for the following reasons: firstly, the pion mass
dependence of $M_\eta$ is very weak. Secondly, corrections to
$M_\mathrm{K}$ appear to be small, as discussed earlier. And thirdly,
only a few ensembles have small $M_\mathrm{PS}\cdot L$, and hence, a
small change in these does not affect the fit result significantly. As
soon as the results for $M_\eta$ get even more precise the analysis
should include finite size corrections to $M_\mathrm{PS}$ and
$M_\mathrm{K}$.

\begin{figure}[t]
 \centering
 \includegraphics[width=.7\linewidth]%
 {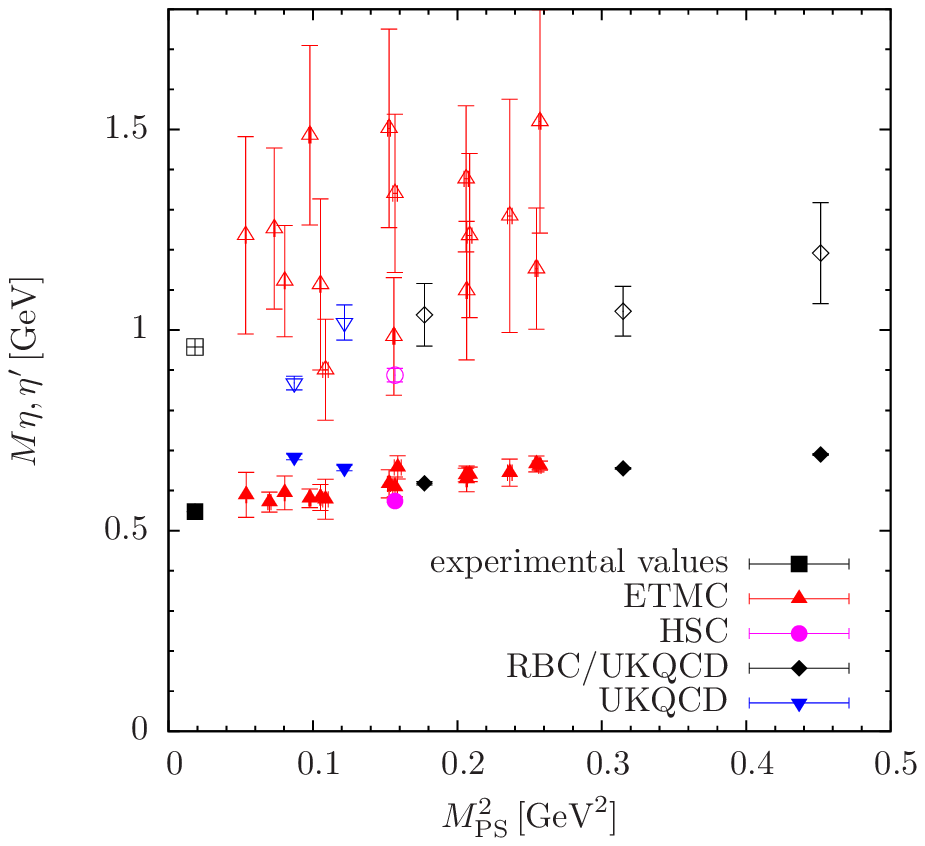}
 \caption{Comparison of our results for $\overline{M}_\eta$ (filled
   symbols) (corrected
   for the mismatch in $M_\mathrm{K}$ as discussed in the text) and
   $M_{\eta'}$ (open symbols) in physical units for all three values of the lattice
   spacing to results from the literature (RBC/UKQCD~\cite{Christ:2010dd},
   HSC~\cite{Dudek:2011tt}, UKQCD~\cite{Gregory:2011sg}. 
   The scale for our points was set using $r_0=0.45(2)\ \mathrm{fm}$.} 
 \label{fig:allmasses}
\end{figure}

In figure~\ref{fig:allmasses} we show a compilation of our results for
$\eta$ masses (corrected for the mismatch in $M_\mathrm{K}$, cf.
figure~\ref{fig:scaling}) and $\eta'$ masses and those available in
the literature for $N_f=2+1$ flavour lattice QCD: in
Ref.~\cite{Kaneko:2009za} $\eta$ and $\eta'$ meson masses have been
computed using $N_f=2+1$ flavours of overlap quarks at one value
of the lattice spacing and large values of the pion mass. In this
reference not enough details are given to be included in our
comparison figure~\ref{fig:allmasses}.
In Ref.~\cite{Christ:2010dd} these masses have been determined
using $N_f=2+1$ flavours of domain wall fermions, again
at a single value of the lattice spacing $a\approx 0.1\
\mathrm{fm}$ and three values of the pion mass ranging from about
$400\ \mathrm{MeV}$ to about $700\ \mathrm{MeV}$. The corresponding
data points are labeled RBC/UKQCD in figure~\ref{fig:allmasses}.
In Ref.~\cite{Dudek:2011tt} the Hadron Spectrum Collaboration (HSC) --
also at a single value of the lattice spacing -- and one value of
the pion mass used Wilson fermions. Finally, in
Ref.~\cite{Gregory:2011sg} data is presented for two values of the
pion mass, each at a different value of the lattice spacing with
staggered fermions, labeled as UKQCD in our plot. As discussed
previously, we used a value of $r_0=0.45(2)\ \mathrm{fm}$ to set the
scale for our data.

Firstly, from figure~\ref{fig:allmasses} it is clear that the results
presented in this paper significantly increase the available data for
$\eta$ and $\eta'$ masses from lattice QCD. In particular, our results
extend to significantly 
lower values of the pion mass than available before. And, within the
statistical uncertainties, all the results from the different lattice
formulations agree, despite the fact that the systematic uncertainties
are not taken into account, and that we used $N_f=2+1+1$ dynamical
quark flavours.

Secondly, figure~\ref{fig:allmasses} allows to compare the statistical
uncertainties quoted by the different collaborations. The figure shows
that our error on the $\eta'$ masses are by far larger than what is
quoted by the other collaborations. RBC/UKQCD~\cite{Christ:2010dd}
have investigated $300$ configurations separated by $20$ HMC
trajectories, and they did not find autocorrelation among these
configurations for the investigated quantities. The
HSC~\cite{Dudek:2011tt} used $479$ configurations also separated by
$20$ HMC trajectories. Both collaborations use a lattice spacing
around $0.12\ \mathrm{fm}$, i.e. significantly larger than our
coarsest value. Compared to these two collaborations we have hence a 
similar number of independent configurations per ensemble, but smaller
values of the lattice spacing. Our method to compute disconnected
contributions involves stochastic noise, which is not the case for
these two collaborations. However, we have tested that the stochastic
noise is not dominantly affecting our results; the gauge noise is
dominant in our data. Compared to RBC/UKQCD
we have a similar number of inversions per independent gauge
configuration, while HSC has many more.  Clearly, the large operator
basis used by HSC will help to extract the states with higher
precision. An explanation for the smaller errors found by RBC/UKQCD
might be the better chiral properties of their formulation, the larger
value of the lattice spacing and the smaller volume in lattice units.
In the staggered investigation by UKQCD~\cite{Gregory:2011sg} a
similar method to ours was used, but with a larger number of
independent configurations. 

We would like to point out again that the large error on $M_{\eta'}$
also reflects the systematic uncertainty in identifying a plateau in
its effective mass, as discussed earlier. 

\subsection[eta and eta' Mixing]
{$\eta$ and $\eta'$ Mixing}

\begin{figure}[t]
 \centering
 \includegraphics[width=.48\linewidth]{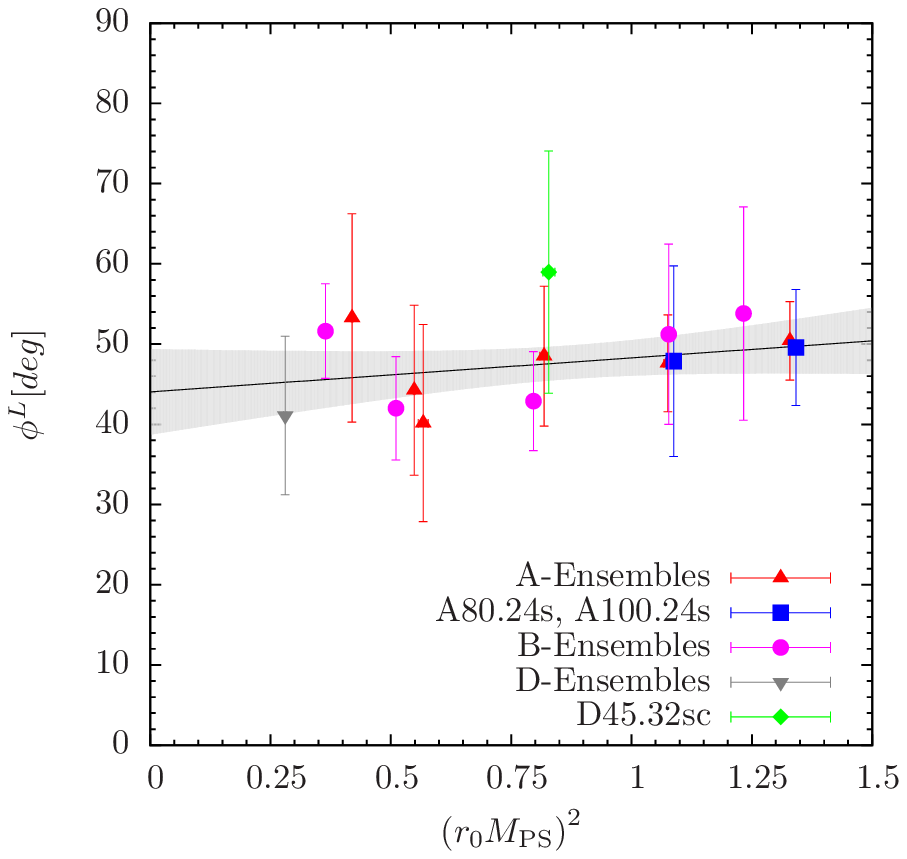}\quad
 \includegraphics[width=.48\linewidth]{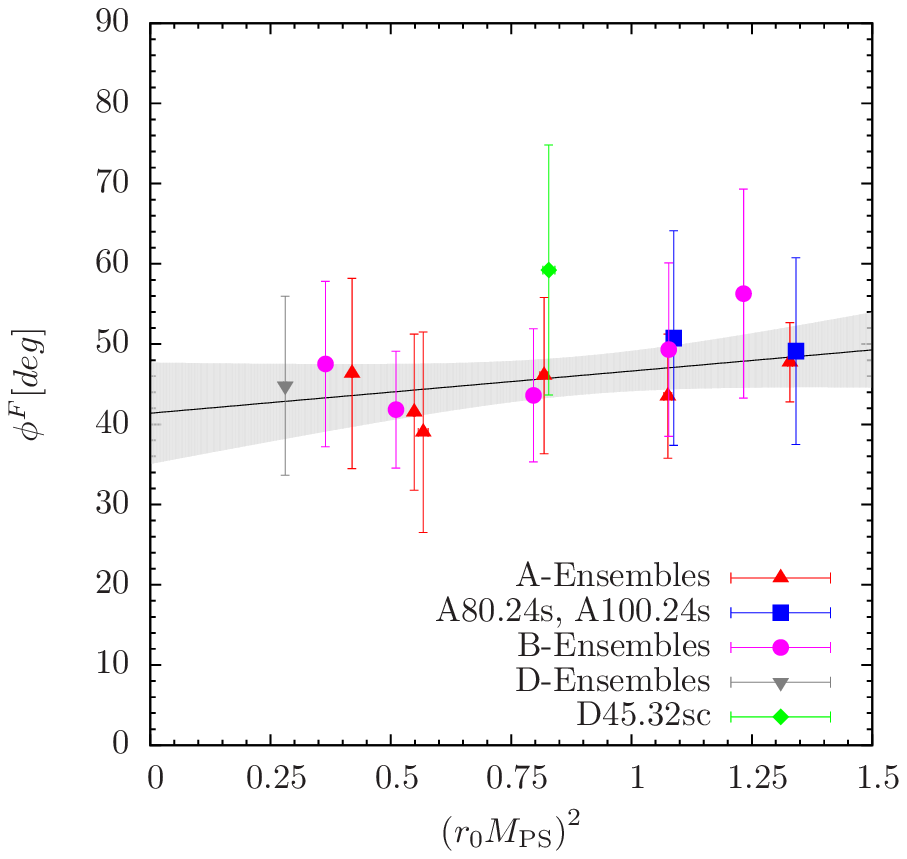}
 \caption{(a) mixing angle $\phi^{L}$ from local
   amplitudes of a $4\times4$-matrix (no charm quark). (b) mixing
   angle $\phi^{F}$ from fuzzed amplitudes of
   $6\times6$-matrix (including charm quark)}
 \label{fig:mixing_angles}
\end{figure}

As mentioned previously, the determination of the $\eta$ and $\eta'$
mixing angles requires the knowledge of the ratio of renormalisation
constants $Z$ as discussed above. 

We determine the amplitudes $A_{\ell,\eta}$, $A_{\ell,\eta'}$,
$A_{s,\eta}$ and $A_{s,\eta'}$ from a factorising fit to the
correlation matrix $\mathcal{C}(t)$ rotated to its original form. For
the factorising fit we consider only light and strange degrees of
freedom (both, local and fuzzed), as the charm does not contribute
within errors. We then determine $\phi_\ell$ and $\phi_s$ from
eq.~(\ref{eq:angles}) and $\phi$ from
eq.~(\ref{eq:meanangle}). The errors are again determined using a
bootstrap procedure with $1000$ bootstrap samples. All results from a
$4\times4$ matrix fit are collected in
tab.~\ref{tab:angles_results}. Results from a $6\times6$ matrix can be
found in the appendix.

The results for $\phi$ are shown in
figure~\ref{fig:mixing_angles}. In the left panel we show
$\phi$ in degrees determined from local amplitudes (L),
and in the right panel from fuzzed amplitudes (F), both as a function of
$(r_0M_\mathrm{PS})^2$. Both agree nicely and we do 
not observe a dependence on the lattice spacing and
$(r_0M_\mathrm{K})^2$, at least within the relatively large
errors. The dependence on $(r_0M_\mathrm{PS})^2$ is weak, the data 
is compatible with a linear extrapolation to the physical pion mass
point, which leads to a value of
\[
\phi = 44(5)^\circ \, ,
\]
using a combined fit to local and fuzzed data. In the octet basis
this would correspond to a $\eta_8,\eta_1$ mixing angle $\theta =\phi
- 54.7^\circ = -10(5)^\circ$. For this extrapolation we ignored a
possible strange and charm quark mass dependence of $\phi$.
Our result for $\phi$ is in
agreement with the results from RBC/UKQCD~\cite{Christ:2010dd},
HSC~\cite{Dudek:2011tt} and an old UKQCD work~\cite{McNeile:2000hf},
which all quote something in between $40^\circ$ and $50^\circ$. In the
recent UKQCD staggered investigation~\cite{Gregory:2011sg} a value of
$34(3)^\circ$ is favoured. Comparing to experimental and
phenomenological
results~\cite{Feldmann:1999uf,DiDonato:2011kr,Gregory:2011sg} we find
excellent  
agreement to results from radiative decays and glueball mixing
($\sim42^\circ$). Results from photon fusion and charm-$\eta$
production favour a value of $\sim 33^\circ$. Note that one has to
keep in mind that our determination of the mixing angle is likely to
be affected by systematic uncertainties.

\begin{figure}[t]
 \centering   
 \includegraphics[width=.48\linewidth]{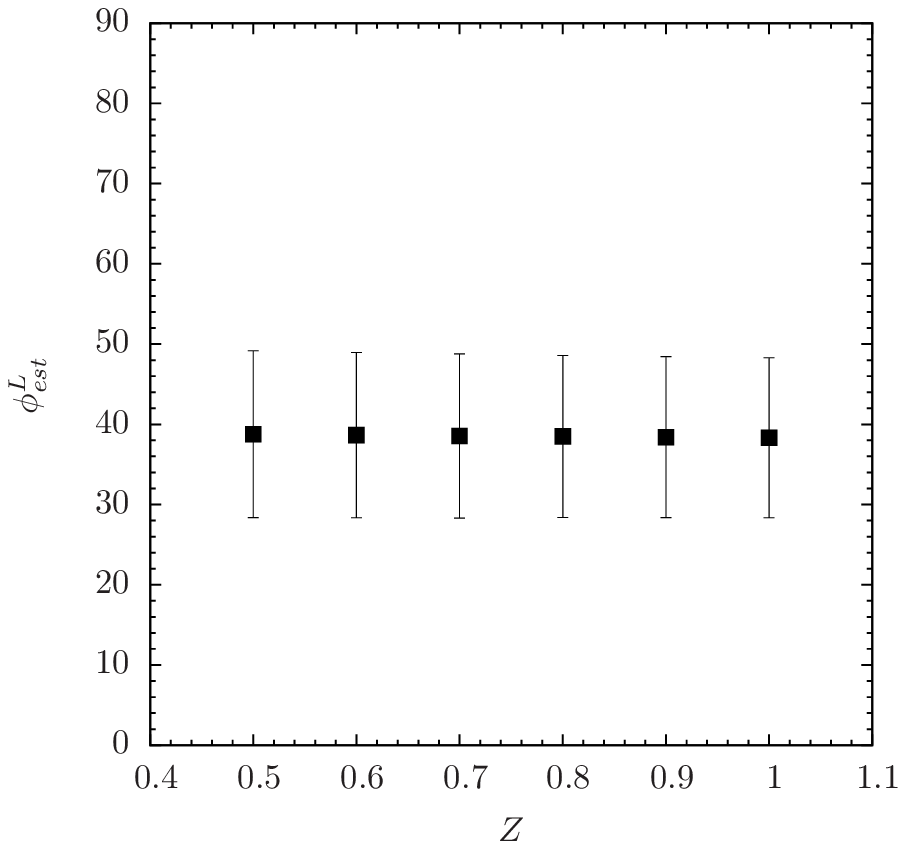}
 \caption{$Z$-dependence of the double-ratio mixing angle
   $\phi^L$ defined in eq.~(\ref{eq:meanangle}) for D15.48 using
   local amplitudes only.} 
 \label{fig:angleZdep}
\end{figure}

We have checked that this determination is not affected by our
uncertainty on the ratio of renormalisation constants $Z$. For 
this we varied $Z$ in a large range from $0.4$ to $1.0$, and we found
that this does not affect the extraction of the mixing angles. As an
example we show the dependence of $\phi^L$ for ensemble $D15.48$
in figure~\ref{fig:angleZdep} on $Z$. The variation introduced by $Z$
is by far smaller than the statistical uncertainty. We observe the
same for the other ensembles and conclude that our evaluation of the
angles is not affected by systematic uncertainties stemming from the
$Z$ ratio within our statistical uncertainties.

While figure~\ref{fig:mixing_angles} indicates that the angle $\phi$
behaves regularly and the linear extrapolation to the physical pion mass
appears to be reasonable, the results for $\phi_\ell$ and $\phi_s$ are
diverse, see table~\ref{tab:angles_results}. We attribute this to
large statistical and systematic uncertainties in these quantities. We
hope to be able to investigate $\phi_\ell$ and $\phi_s$ further once
we have improved our determination of the $\eta'$ state.


\section{Summary and Discussion}

We presented the first computation of $\eta$ and $\eta'$ meson masses
from lattice QCD with degenerate up/down, heavier strange and even
heavier charm dynamical quarks in the Wilson twisted mass
formulation. Our results based on ETMC gauge configurations cover
three values of the lattice spacing from $\sim0.09\ \mathrm{fm}$ down
to $\sim 0.06\ \mathrm{fm}$ and a range of pion masses from $230\
\mathrm{MeV}$ to $500\ \mathrm{MeV}$. The results presented here are
therefore the first covering three values of the lattice spacing and a
range of pion masses down to $230\ \mathrm{MeV}$. Kaon and D-meson
masses are close to their physical values. We observe in both, $\eta$
and $\eta'$ masses, only a mild dependence on the light quark 
mass. The $\eta$ meson mass can be determined with high statistical
accuracy, while the $\eta'$ suffers from noise and large
autocorrelation times.

In our simulations, the bare strange and charm quark masses were kept
fixed for each value of the lattice spacing separately, apart from two
additional $A$-ensembles. As the kaon mass values deviate from the
experimental values by up to 10\%, we applied three methods to correct
for this mismatch: firstly, by estimating the strange quark mass
dependence from the aforementioned additional $A$-ensembles, secondly
by considering the ratio $M_\eta/M_\mathrm{K}$ and thirdly, by using
the GMO relation eq.~\ref{eq:GMOmasses}. All three methods seem to
indicate that lattice artifacts are only weakly affecting the
$\eta$ mass. As our best estimate we extract 
\[
M_\eta = 557(15)_\mathrm{stat}(45)_\mathrm{sys}\ \mathrm{MeV}
\]
by a weighted average over the three methods. The first error is
statistical and the second systematic from the continuum
extrapolation. The result is in good agreement with the experimental
value for $M_\eta$.

We also determine the $\eta$ and $\eta'$ mixing angle $\phi$ using
amplitudes from a factorising fit model. Again, we observe little
light quark mass and lattice spacing dependence within errors. But
this time also the strange quark mass dependence is smaller than our
(large) statistical uncertainty. Our best estimate for the mixing
angle $\phi$ is 
\[
\phi = 44(5)^\circ
\] 
with statistical error only. In an octet basis this corresponds to
$\theta = -10(5)^\circ$. The agreement of our estimate for $\phi$ with
most experimental and
phenomenological~\cite{Feldmann:1999uf,DiDonato:2011kr} and most other
lattice
determinations~\cite{Christ:2010dd,Dudek:2011tt,McNeile:2000hf} is
excellent. The value is also very close to the quadratic GMO estimate
of $44.7^\circ$. 
But results from photon fusion and charm-$\eta$ production,
as well as the lattice determination from ref.~\cite{Gregory:2011sg}
find a value of $\sim 34^\circ$.
Estimating the systematics for the angle becomes
difficult, because the statistical uncertainties for each ensemble are
large. 


It is evident that the $\eta'$ mass is affected by large statistical
and systematical uncertainties. It also shows a large autocorrelation
time. Hence, we currently cannot make a reliable
estimate of the physical $\eta'$ mass value. We therefore plan in the 
future to apply additional noise reduction techniques to get a better
signal also for the $\eta'$ state. There are actually two promising
approaches, the point-to-point method described in
ref.~\cite{Jansen:2008wv} and an extension of the noise reduction
trick used here only for estimating the light disconnected
contributions. Also, investigating the SVD solver further might be a
promising way forward.

Finally, determining the flavour singlet decay constants is of large
phenomenological interest. We are currently investigating our signals
for $\langle0|A_\mu|\eta\rangle$ and $\langle0|A_\mu|\eta'\rangle$.

\subsection*{Acknowledgements} 

We thank J.~Daldrop, E.~Gregory, K.~Jansen, B.~Kubis, C.~McNeile,
U.-G.~Mei{\ss}ner, M.~Petschlies, 
M.~Wagner and F.~Zimmermann for useful discussions. We thank U.~Wenger 
for his help in determining the Sommer parameter. We are indebted to
G.~Herdoiza for a careful reading of the manuscript and very useful
comments. C.U. would like to thank M.~Mahoney for a careful reading of
the manuscript. We thank the
members of ETMC for the most enjoyable collaboration. The computer
time for this project was made available to us by the John von
Neumann-Institute for Computing (NIC) on the JUDGE and Jugene systems
in J{\"u}lich and the IDRIS (CNRS) computing center in Orsay. In
particular we thank U.-G.~Mei{\ss}ner for granting us 
access on JUDGE. This project was funded by the DFG as a project in
the SFB/TR 16. Two of the authors (K. O. and C.U.) were supported by
the Bonn-Cologne Graduate School (BCGS) of Physics and Astronomie. The
open source software packages tmLQCD~\cite{Jansen:2009xp},
Lemon~\cite{Deuzeman:2011wz} and R~\cite{R:2005} have been used. 

\bibliographystyle{h-physrev5}
\bibliography{bibliography}
\begin{appendix}
  \section{Analysis Details}
\label{sec:app}

In this appendix we summarise the technical details of our numerical
methods for extracting masses and angles.

\subsection{Generalised Eigenvalue Problem (GEVP)}

\begin{table}[t!]
 \centering
 \begin{tabular*}{.8\textwidth}{@{\extracolsep{\fill}}lcccccccc}
  \hline\hline
  ensemble & $t_{0}^{\eta}$ & $t_{1}^{\eta}$ & $t_{2}^{\eta}$ &
  $\left(\chi^{2}\right)^{\eta}/\mathrm{dof}$ & $t_{0}^{\eta'}$ &
  $t_{1}^{\eta'}$ & $t_{2}^{\eta'}$ &
  $\left(\chi^{2}\right)^{\eta'}/\mathrm{dof}$ \\ 
  \hline\hline
  A30.32   & 2 & 7 & 17 & 0.205 & 1 & 2 & 11 & 0.277 \\
  A40.24   & 2 & 7 & 16 & 0.135 & 1 & 2 & 10 & 0.185 \\
  A40.32   & 2 & 7 & 15 & 0.092 & 1 & 2 & 10 & 0.130 \\
  A60.24   & 2 & 7 & 15 & 0.137 & 1 & 2 & 10 & 0.296 \\
  A80.24   & 2 & 7 & 16 & 0.108 & 1 & 2 & 10 & 0.137 \\
  A100.24  & 2 & 7 & 15 & 0.214 & 1 & 2 & 11 & 0.137 \\
  \hline
  A80.24s  & 2 & 7 & 17 & 0.230 & 1 & 2 & 11 & 0.440 \\
  A100.24s & 2 & 7 & 16 & 0.086 & 1 & 2 & 10 & 0.116 \\
  \hline
  B25.32   & 3 & 6 & 11 & 0.222 & 1 & 2 & 11 & 0.463 \\
  B35.32   & 2 & 8 & 17 & 0.110 & 1 & 2 & 10 & 0.433 \\
  B55.32   & 2 & 8 & 18 & 0.167 & 1 & 2 & 11 & 0.366 \\
  B75.32   & 2 & 8 & 14 & 0.301 & 1 & 2 & 10 & 0.338 \\
  B85.24   & 2 & 8 & 16 & 0.106 & 1 & 2 & 10 & 0.127 \\
  \hline
  D15.48   & 3 & 8 & 16 & 0.190 & 2 & 3 & 12 & 0.114 \\
  D45.32sc & 3 & 8 & 19 & 0.1743 & 2 & 3 & 18 & 0.105 \\
  \hline\hline
  \vspace*{0.1cm}
 \end{tabular*}
 \caption{Parameters of the GEVP applied to
   the $6\times6$-matrix from local and fuzzed operators.} 
 \label{tab:gevp_params}
\end{table}

For the determination of masses and identification of flavour contents
of the states we employ the variational approach in
eq.~(\ref{eq:gevp}). We extract the eigenvalues $\lambda^{(n)}(t,t_0)$
from the GEVP and determine then the masses from a fit to
$\lambda$. The errors of $\lambda$ are determined using a bootstrap
procedure with $1000$ bootstrap samples, after the data is blocked in
order to account for autocorrelation.
An overview of the remaining GEVP and fitting parameters is listed
in table \ref{tab:gevp_params}. Due to the rather large difference in
the signal-to-noise ratio between the two lowest lying states we
use two different approaches for $\eta$ and $\eta'$: 
\begin{enumerate}
\item For the ground-state ($\eta$) we fit a single $\cosh$ in $t$
  in a region $\left[t_{1}^{\eta},t_{2}^{\eta}\right]$ to our data for
  $\lambda^\eta(t,t_0)$. The
  lower bound of the fit-range $t_1^{\eta}$ is chosen by visual
  inspection of the effective mass plot to lie at the beginning of the plateau and
  also such that further increasing $t_1^{\eta}$ does not change the
  value of the resulting mass within errors. The latter is also true
  for the choice of the starting value $t_0^{\eta}$ in
  eq.~(\ref{eq:gevp}). In principal choosing $t_0$ larger leads to
  smaller masses, though due to noise we are only able to moderately
  increase $t_0$ before the error gets too large. 

\item For the first excited state ($\eta'$) we perform a three state
  $\cosh$-fit to the data of $\lambda^{\eta'}(t,t_0)$, starting from
  the lowest $t_1^{\eta'}$ possible, 
  i.e. $t_1^{\eta'}=t_0^{\eta'}+1$ in order to use as many points as
  possible. This is necessary because for many ensembles there is no
  clear plateau in the effective masses reached before the signal is
  lost in noise. Therefore, 
  this procedure is a major source of systematic error for the 
  determination of $\eta'$ masses, at least for those cases where only
  few points are available. Only for the $D$-Ensembles it turns out to
  be possible to choose $t_0^{\eta'}>1$, for all other ensembles we
  had to use $t_0^{\eta'}=1$. 
\end{enumerate}
The upper bound of the fit range $t_{2}^{n}$ is independently
determined for every state $n$ by the last eigenvalue
$\lambda^{\left(n\right)}\left(t_2, t_0\right)$ distinguishable from
noise. Note that the value of $t_2$ is not very important for the fit,
as eigenvalues at large $t$ have typically large errors and therefore
do not contribute much to the fit.

\subsection{Factorising fit model}
For the extraction of amplitudes which are required to calculate mixing
angles we employ the factorising fit model as detailed in
eq.~(\ref{eq:fit}) to the matrix in the original twisted basis in
eq.~(\ref{eq:corrmatrix}). We limit ourselves to $n=2$ in
eq.~(\ref{eq:fit}), i.e. a two state fit. As discussed before the
ratio of renormalisation constant $Z=Z_P/Z_S$ does not affect
masses and angles, but only amplitudes. For the results of the angles
we used the values summarised in table~\ref{tab:Zratios}. They have
been obtained by matching a mixed action to the unitary action and
extrapolating to the chiral limit, see
ref.~\cite{Farchioni:2010tb}. These values agree well with the RI-MOM
determination of ETMC~\cite{ETM:2011aa,ETM:rimom}, and the method we
discussed in the text. 

\begin{table}[t!]
 \centering
 \begin{tabular*}{.8\textwidth}{@{\extracolsep{\fill}}lccc}
  \hline\hline
  $\beta$   & 1.90    & 1.95    & 2.10 \\
  $Z$ & 0.6703(8) & 0.6859(9) & 0.7493(11) \\
  \hline\hline
  \vspace*{0.1cm}
 \end{tabular*}
 \caption{Values for $Z$ from matching mixed action
   (Osterwalder-Seiler) to unitary approach.}
 \label{tab:Zratios}
\end{table}

As we have already determined the masses from the GEVP, we use those
together with their errors (see table
\ref{tab:masses_results}) as priors to our factorising fit. This
stabilises the fit procedure, but we have checked that it does not
affect the result. We always use an uncorrelated $\chi^2$ function,
because the correlation matrix is too noisy. Table
\ref{tab:factorizing_params} lists the input parameters and the
resulting uncorrelated $\chi^2 / \mathrm{dof}$ for the fits to the
$4\times4$ (light and strange degrees of freedom, local and fuzzed)
and $6\times6$ (light, strange charm, and local and fuzzed) correlator
matrix. The results quoted in the main text for the angles were
determined using the $4\times4$ matrix, because the charm does not
contribute. However, with the $6\times6$ matrix we obtain almost
identical results.

The value of the lower bound of the fit range $t_1$ is chosen to be
constant for every value of the lattice spacing, whereas the choice of
the upper bound $t_2$ is determined by the requirement that $\chi^2 /
\mathrm{dof}<1$, and that it is close to the $t_2^\eta$ used for
GEVP. 

\begin{table}[t!]
 \centering
 \begin{tabular*}{.8\textwidth}{@{\extracolsep{\fill}}lcccc}
  \hline\hline
  ensemble & $t_1$ & $t_2$ & $\chi^2_{4\times4}$ & $\chi^2_{6\times6}$ \\
  \hline\hline
  A30.32   & 6 & 18 & 0.252 & 0.561 \\
  A40.24   & 6 & 14 & 0.144 & 0.626 \\
  A40.32   & 6 & 15 & 0.466 & 0.628 \\
  A60.24   & 6 & 14 & 0.229 & 0.470 \\
  A80.24   & 6 & 14 & 0.204 & 0.545 \\
  A100.24  & 6 & 13 & 0.165 & 0.395 \\
  \hline
  A80.24s  & 6 & 17 & 0.803 & 0.986 \\
  A100.24s & 6 & 13 & 0.063 & 0.293 \\
  \hline
  B25.32   & 6 & 16 & 0.326 & 0.857 \\
  B35.32   & 6 & 15 & 0.283 & 0.694 \\
  B55.32   & 6 & 16 & 0.836 & 0.774 \\
  B75.32   & 6 & 12 & 0.449 & 0.636 \\
  B85.24   & 6 & 14 & 0.172 & 0.469 \\
  \hline
  D15.48   & 7 & 17 & 0.328 & 0.567 \\
  D45.32sc & 7 & 15 & 0.388 & 0.686 \\
  \hline\hline
  \vspace*{0.1cm}
 \end{tabular*}
 \caption{Parameters for the factorising fit.}
 \label{tab:factorizing_params}
\end{table}

\end{appendix}
\end{document}